\documentclass[twocolumn]{aastex631}

\usepackage{amsmath, xparse}

\journalinfo{}

\begin{document}

\title{Hydrogen Epoch of Reionization Array (HERA) Phase II Deployment and Commissioning}

\author[0000-0002-2293-9639]{Lindsay M. Berkhout}
\affiliation{School of Earth and Space Exploration, Arizona State University, Tempe, AZ}

\correspondingauthor{Lindsay M. Berkhout}
\email{lmberkhout@asu.edu}

\author[0000-0002-0917-2269]{Daniel C. Jacobs}
\affiliation{School of Earth and Space Exploration, Arizona State University, Tempe, AZ}

\author{Zuhra Abdurashidova}
\affiliation{Department of Astronomy, University of California, Berkeley, CA}

\author{Tyrone  Adams}
\affiliation{South African Radio Astronomy Observatory, Black River Park, 2 Fir Street, Observatory, Cape Town, 7925, South Africa}

\author[0000-0002-4810-666X]{James E. Aguirre}
\affiliation{Department of Physics and Astronomy, University of Pennsylvania, Philadelphia, PA}

\author{Paul  Alexander}
\affiliation{Cavendish Astrophysics, University of Cambridge, Cambridge, UK}

\author{Zaki S. Ali}
\affiliation{Department of Astronomy, University of California, Berkeley, CA}

\author{Rushelle  Baartman}
\affiliation{South African Radio Astronomy Observatory, Black River Park, 2 Fir Street, Observatory, Cape Town, 7925, South Africa}

\author{Yanga  Balfour}
\affiliation{South African Radio Astronomy Observatory, Black River Park, 2 Fir Street, Observatory, Cape Town, 7925, South Africa}

\author[0000-0001-9428-8233]{Adam P. Beardsley}
\affiliation{School of Earth and Space Exploration, Arizona State University, Tempe, AZ}
\affiliation{Department of Physics, Winona State University, Winona, MN}

\author[0000-0002-0916-7443]{Gianni  Bernardi}
\affiliation{INAF-Istituto di Radioastronomia, via Gobetti 101, 40129 Bologna, Italy}
\affiliation{Department of Physics and Electronics, Rhodes University, PO Box 94, Grahamstown, 6140, South Africa}
\affiliation{South African Radio Astronomy Observatory, Black River Park, 2 Fir Street, Observatory, Cape Town, 7925, South Africa}

\author{Tashalee S. Billings}
\affiliation{Department of Physics and Astronomy, University of Pennsylvania, Philadelphia, PA}

\author[0000-0002-8475-2036]{Judd D. Bowman}
\affiliation{School of Earth and Space Exploration, Arizona State University, Tempe, AZ}

\author{Richard F. Bradley}
\affiliation{National Radio Astronomy Observatory, Charlottesville, VA}

\author[0000-0001-5668-3101]{Philip  Bull}
\affiliation{Jodrell Bank Centre for Astrophysics, University of Manchester, Manchester, M13 9PL, United Kingdom}
\affiliation{Department of Physics and Astronomy,  University of Western Cape, Cape Town, 7535, South Africa}

\author[0000-0002-8465-9341]{Jacob  Burba}
\affiliation{Jodrell Bank Centre for Astrophysics, University of Manchester, Manchester, M13 9PL, United Kingdom}

\author{Steven  Carey}
\affiliation{Cavendish Astrophysics, University of Cambridge, Cambridge, UK}

\author[0000-0001-6647-3861]{Chris L. Carilli}
\affiliation{National Radio Astronomy Observatory, Socorro, NM 87801, USA}

\author[0000-0002-3839-0230]{Kai-Feng Chen}
\affiliation{MIT Kavli Institute, Massachusetts Institute of Technology, Cambridge, MA}
\affiliation{Department of Physics, Massachusetts Institute of Technology, Cambridge, MA}

\author{Carina  Cheng}
\affiliation{Department of Astronomy, University of California, Berkeley, CA}

\author{Samir  Choudhuri}
\affiliation{Queen Mary University London, London E1 4NS, UK}
\affiliation{Centre for Strings, Gravitation and Cosmology, Department of Physics, Indian Institute of Technology Madras, Chennai 600036, India}

\author[0000-0003-3197-2294]{David R. DeBoer}
\affiliation{Radio Astronomy Lab, University of California, Berkeley, CA}

\author{Eloy  de~Lera~Acedo}
\affiliation{Cavendish Astrophysics, University of Cambridge, Cambridge, UK}

\author{Matt  Dexter}
\affiliation{Radio Astronomy Lab, University of California, Berkeley, CA}

\author[0000-0003-3336-9958]{Joshua S. Dillon}
\affiliation{Department of Astronomy, University of California, Berkeley, CA}

\author{Scott  Dynes}
\affiliation{MIT Kavli Institute, Massachusetts Institute of Technology, Cambridge, MA}

\author{Nico  Eksteen}
\affiliation{South African Radio Astronomy Observatory, Black River Park, 2 Fir Street, Observatory, Cape Town, 7925, South Africa}

\author{John  Ely}
\affiliation{Cavendish Astrophysics, University of Cambridge, Cambridge, UK}

\author[0000-0002-0086-7363]{Aaron  Ewall-Wice}
\affiliation{Department of Astronomy, University of California, Berkeley, CA}
\affiliation{Department of Physics, University of California, Berkeley, CA}

\author{Nicolas  Fagnoni}
\affiliation{Cavendish Astrophysics, University of Cambridge, Cambridge, UK}

\author{Randall  Fritz}
\affiliation{South African Radio Astronomy Observatory, Black River Park, 2 Fir Street, Observatory, Cape Town, 7925, South Africa}

\author[0000-0002-0658-1243]{Steven R. Furlanetto}
\affiliation{Department of Physics and Astronomy, University of California, Los Angeles, CA}

\author{Kingsley  Gale-Sides}
\affiliation{Cavendish Astrophysics, University of Cambridge, Cambridge, UK}

\author{Hugh  Garsden}
\affiliation{Queen Mary University London, London E1 4NS, UK}
\affiliation{Jodrell Bank Centre for Astrophysics, University of Manchester, Manchester, M13 9PL, United Kingdom}

\author{Bharat Kumar Gehlot}
\affiliation{School of Earth and Space Exploration, Arizona State University, Tempe, AZ}

\author{Abhik  Ghosh}
\affiliation{Department of Physics and Astronomy,  University of Western Cape, Cape Town, 7535, South Africa}

\author{Brian  Glendenning}
\affiliation{National Radio Astronomy Observatory, Socorro, NM}

\author{Adelie  Gorce}
\affiliation{Department of Physics and McGill Space Institute, McGill University, 3600 University Street, Montreal, QC H3A 2T8, Canada}

\author[0000-0002-0829-167X]{Deepthi  Gorthi}
\affiliation{Department of Astronomy, University of California, Berkeley, CA}

\author[0000-0002-4085-2094]{Bradley  Greig}
\affiliation{School of Physics, University of Melbourne, Parkville, VIC 3010, Australia}

\author{Jasper  Grobbelaar}
\affiliation{South African Radio Astronomy Observatory, Black River Park, 2 Fir Street, Observatory, Cape Town, 7925, South Africa}

\author{Ziyaad  Halday}
\affiliation{South African Radio Astronomy Observatory, Black River Park, 2 Fir Street, Observatory, Cape Town, 7925, South Africa}

\author[0000-0001-7532-645X]{Bryna J. Hazelton}
\affiliation{Department of Physics, University of Washington, Seattle, WA}
\affiliation{eScience Institute, University of Washington, Seattle, WA}

\author[0000-0002-4117-570X]{Jacqueline N. Hewitt}
\affiliation{MIT Kavli Institute, Massachusetts Institute of Technology, Cambridge, MA}
\affiliation{Department of Physics, Massachusetts Institute of Technology, Cambridge, MA}

\author{Jack  Hickish}
\affiliation{Radio Astronomy Lab, University of California, Berkeley, CA}

\author{Tian  Huang}
\affiliation{Cavendish Astrophysics, University of Cambridge, Cambridge, UK}

\author{Alec  Josaitis}
\affiliation{Cavendish Astrophysics, University of Cambridge, Cambridge, UK}

\author{Austin  Julius}
\affiliation{South African Radio Astronomy Observatory, Black River Park, 2 Fir Street, Observatory, Cape Town, 7925, South Africa}

\author{MacCalvin  Kariseb}
\affiliation{South African Radio Astronomy Observatory, Black River Park, 2 Fir Street, Observatory, Cape Town, 7925, South Africa}

\author[0000-0002-8211-1892]{Nicholas S. Kern}
\affiliation{MIT Kavli Institute, Massachusetts Institute of Technology, Cambridge, MA}
\affiliation{Department of Physics, Massachusetts Institute of Technology, Cambridge, MA}
\affiliation{NASA Hubble Fellow}

\author[0000-0002-1876-272X]{Joshua  Kerrigan}
\affiliation{Department of Physics, Brown University, Providence, RI}

\author[0000-0001-5421-8927]{Honggeun Kim}
\affiliation{MIT Kavli Institute, Massachusetts Institute of Technology, Cambridge, MA}
\affiliation{Department of Physics, Massachusetts Institute of Technology, Cambridge, MA}

\author[0000-0003-0953-313X]{Piyanat  Kittiwisit}
\affiliation{Department of Physics and Astronomy,  University of Western Cape, Cape Town, 7535, South Africa}

\author[0000-0001-6744-5328]{Saul A. Kohn}
\affiliation{Department of Physics and Astronomy, University of Pennsylvania, Philadelphia, PA}

\author[0000-0002-2950-2974]{Matthew  Kolopanis}
\affiliation{School of Earth and Space Exploration, Arizona State University, Tempe, AZ}

\author{Adam  Lanman}
\affiliation{Department of Physics, Brown University, Providence, RI}

\author[0000-0002-4693-0102]{Paul  La~Plante}
\affiliation{Department of Astronomy, University of California, Berkeley, CA}
\affiliation{Department of Physics and Astronomy, University of Pennsylvania, Philadelphia, PA}
\affiliation{Department of Computer Science, University of Nevada, Las Vegas, NV}
\affiliation{Nevada Center for Astrophysics, University of Nevada, Las Vegas, NV}

\author[0000-0001-6876-0928]{Adrian  Liu}
\affiliation{Department of Physics and McGill Space Institute, McGill University, 3600 University Street, Montreal, QC H3A 2T8, Canada}

\author{Anita  Loots}
\affiliation{South African Radio Astronomy Observatory, Black River Park, 2 Fir Street, Observatory, Cape Town, 7925, South Africa}

\author[0000-0001-8108-0986]{Yin-Zhe  Ma}
\affiliation{Department of Physics, Stellenbosch University, Matieland 7602, South Africa}

\author{David Harold~Edward MacMahon}
\affiliation{Radio Astronomy Lab, University of California, Berkeley, CA}

\author{Lourence  Malan}
\affiliation{South African Radio Astronomy Observatory, Black River Park, 2 Fir Street, Observatory, Cape Town, 7925, South Africa}

\author{Cresshim  Malgas}
\affiliation{South African Radio Astronomy Observatory, Black River Park, 2 Fir Street, Observatory, Cape Town, 7925, South Africa}

\author{Keith  Malgas}
\affiliation{South African Radio Astronomy Observatory, Black River Park, 2 Fir Street, Observatory, Cape Town, 7925, South Africa}

\author{Bradley  Marero}
\affiliation{South African Radio Astronomy Observatory, Black River Park, 2 Fir Street, Observatory, Cape Town, 7925, South Africa}

\author{Zachary E. Martinot}
\affiliation{Department of Physics and Astronomy, University of Pennsylvania, Philadelphia, PA}

\author[0000-0003-3374-1772]{Andrei  Mesinger}
\affiliation{Scuola Normale Superiore, 56126 Pisa, PI, Italy}

\author{Mathakane  Molewa}
\affiliation{South African Radio Astronomy Observatory, Black River Park, 2 Fir Street, Observatory, Cape Town, 7925, South Africa}

\author[0000-0001-7694-4030]{Miguel F. Morales}
\affiliation{Department of Physics, University of Washington, Seattle, WA}

\author{Tshegofalang  Mosiane}
\affiliation{South African Radio Astronomy Observatory, Black River Park, 2 Fir Street, Observatory, Cape Town, 7925, South Africa}

\author[0000-0003-3059-3823]{Steven G. Murray}
\affiliation{Scuola Normale Superiore, 56126 Pisa, PI, Italy}
\affiliation{School of Earth and Space Exploration, Arizona State University, Tempe, AZ}

\author[0000-0001-7776-7240]{Abraham R. Neben}
\affiliation{MIT Kavli Institute, Massachusetts Institute of Technology, Cambridge, MA}
\affiliation{Department of Physics, Massachusetts Institute of Technology, Cambridge, MA}

\author{Bojan  Nikolic}
\affiliation{Cavendish Astrophysics, University of Cambridge, Cambridge, UK}

\author[0000-0002-5445-6586]{Chuneeta Devi Nunhokee}
\affiliation{International Centre for Radio Astronomy Research, Curtin University, Bentley, WA 6102, Australia}
\affiliation{ARC Centre of Excellence for All Sky Astrophysics in 3 Dimensions (ASTRO 3D), Bentley, WA 6102, Australia}

\author{Hans  Nuwegeld}
\affiliation{South African Radio Astronomy Observatory, Black River Park, 2 Fir Street, Observatory, Cape Town, 7925, South Africa}

\author[0000-0002-5400-8097]{Aaron R. Parsons}
\affiliation{Department of Astronomy, University of California, Berkeley, CA}

\author[0000-0003-0073-5528]{Robert  Pascua}
\affiliation{Department of Astronomy, University of California, Berkeley, CA}
\affiliation{Department of Physics and McGill Space Institute, McGill University, 3600 University Street, Montreal, QC H3A 2T8, Canada}

\author[0000-0002-9457-1941]{Nipanjana  Patra}
\affiliation{Department of Astronomy, University of California, Berkeley, CA}

\author{Samantha  Pieterse}
\affiliation{South African Radio Astronomy Observatory, Black River Park, 2 Fir Street, Observatory, Cape Town, 7925, South Africa}

\author{Yuxiang  Qin}
\affiliation{School of Physics, University of Melbourne, Parkville, VIC 3010 Australia}
\affiliation{Scuola Normale Superiore, 56126 Pisa, PI, Italy}

\author{Eleanor Rath}
\affiliation{MIT Kavli Institute, Massachusetts Institute of Technology, Cambridge, MA}
\affiliation{Department of Physics, Massachusetts Institute of Technology, Cambridge, MA}

\author{Nima  Razavi-Ghods}
\affiliation{Cavendish Astrophysics, University of Cambridge, Cambridge, UK}

\author{Daniel  Riley}
\affiliation{MIT Kavli Institute, Massachusetts Institute of Technology, Cambridge, MA}

\author{James  Robnett}
\affiliation{National Radio Astronomy Observatory, Socorro, NM 87801, USA}

\author{Kathryn  Rosie}
\affiliation{South African Radio Astronomy Observatory, Black River Park, 2 Fir Street, Observatory, Cape Town, 7925, South Africa}

\author{Mario G. Santos}
\affiliation{South African Radio Astronomy Observatory, Black River Park, 2 Fir Street, Observatory, Cape Town, 7925, South Africa}
\affiliation{Department of Physics and Astronomy,  University of Western Cape, Cape Town, 7535, South Africa}

\author[0000-0002-2871-0413]{Peter  Sims}
\affiliation{Department of Physics and McGill Space Institute, McGill University, 3600 University Street, Montreal, QC H3A 2T8, Canada}

\author[0000-0001-7755-902X]{Saurabh  Singh}
\affiliation{Department of Physics and McGill Space Institute, McGill University, 3600 University Street, Montreal, QC H3A 2T8, Canada}
\affiliation{Raman Research Institute}

\author{Dara  Storer}
\affiliation{Department of Physics, University of Washington, Seattle, WA}

\author{Hilton  Swarts}
\affiliation{South African Radio Astronomy Observatory, Black River Park, 2 Fir Street, Observatory, Cape Town, 7925, South Africa}

\author{Jianrong  Tan}
\affiliation{Department of Physics and Astronomy, University of Pennsylvania, Philadelphia, PA}

\author[0000-0003-1602-7868]{Nithyanandan  Thyagarajan}
\affiliation{Commonwealth Scientific and Industrial Research Organisation (CSIRO), Space \& Astronomy, P. O. Box 1130, Bentley, WA 6102, Australia}
\affiliation{National Radio Astronomy Observatory, Socorro, NM 87801, USA}

\author{Pieter  van~Wyngaarden}
\affiliation{South African Radio Astronomy Observatory, Black River Park, 2 Fir Street, Observatory, Cape Town, 7925, South Africa}

\author[0000-0003-3734-3587]{Peter K.~G. Williams}
\affiliation{Center for Astrophysics, Harvard \& Smithsonian, Cambridge, MA}
\affiliation{American Astronomical Society, Washington, DC}

\author{Haoxuan  Zheng}
\affiliation{Department of Physics, Massachusetts Institute of Technology, Cambridge, MA}

\author[0000-0001-5112-2567]{Zhilei Xu}
\affiliation{MIT Kavli Institute, Massachusetts Institute of Technology, Cambridge, MA}
 
\begin{abstract}
This paper presents the design and deployment of the Hydrogen Epoch of Reionization Array (HERA) phase II system. HERA is designed as a staged experiment targeting 21 cm emission measurements of the Epoch of Reionization. First results from the phase I array are published as of early 2022, and deployment of the phase II system is nearing completion. We describe the design of the phase II system and discuss progress on commissioning and future upgrades. As HERA is a designated Square Kilometer Array (SKA) pathfinder instrument, we also show a number of ``case studies'' that investigate systematics seen while commissioning the phase II system, which may be of use in the design and operation of future arrays. Common pathologies are likely to manifest in similar ways across instruments, and many of these sources of contamination can be mitigated once the source is identified.

\end{abstract}

\section{Introduction} \label{sec:intro}
The Hydrogen Epoch of Reionization Array (HERA, \citealp{DeBoer_2017}) is a staged experiment targeting precision measurements of the 21 cm Hydrogen line during the formation of the first stars in the Cosmic Dawn and the subsequent  Epoch of Reionization (EOR) when the intergalactic medium (IGM) was ionized. HERA targets the redshifted 21 cm hyperfine transition line of neutral hydrogen. In this paper we describe the as-built design of the HERA instrument and discuss lessons learned in the process.

\begin{figure*}[ht]
    \centering
    \includegraphics[width=2\columnwidth]{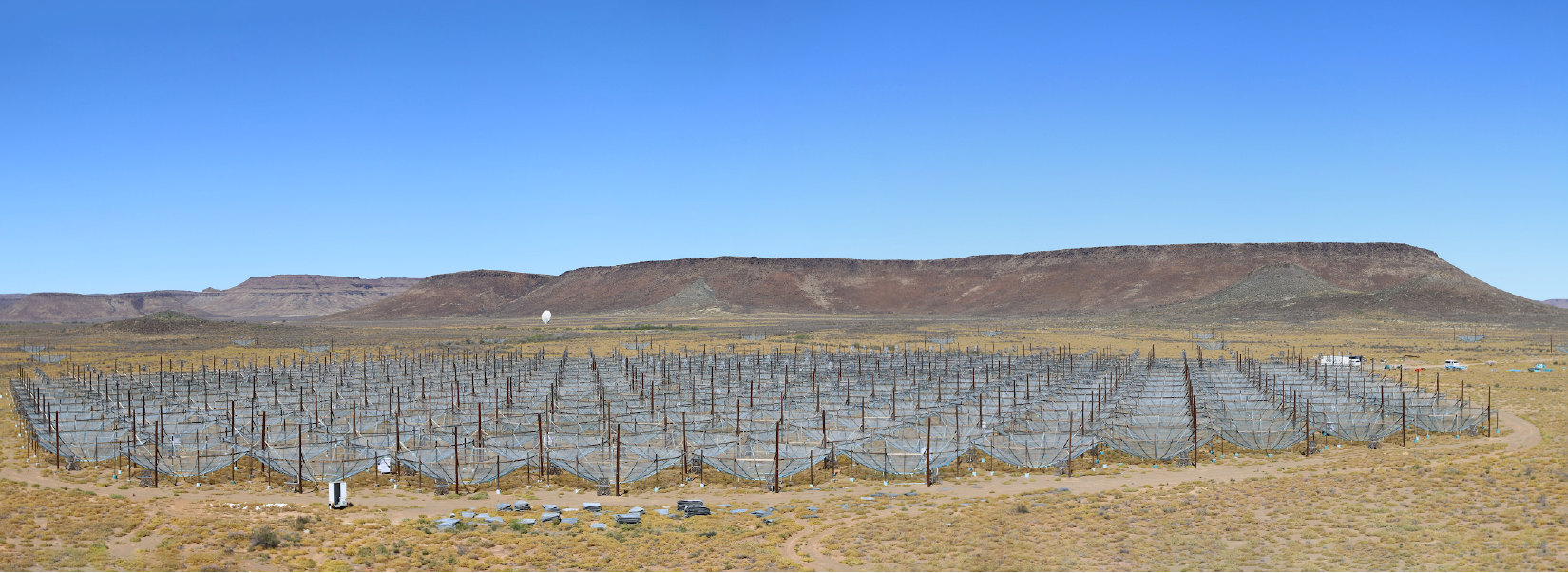}
    \includegraphics[width=2\columnwidth]{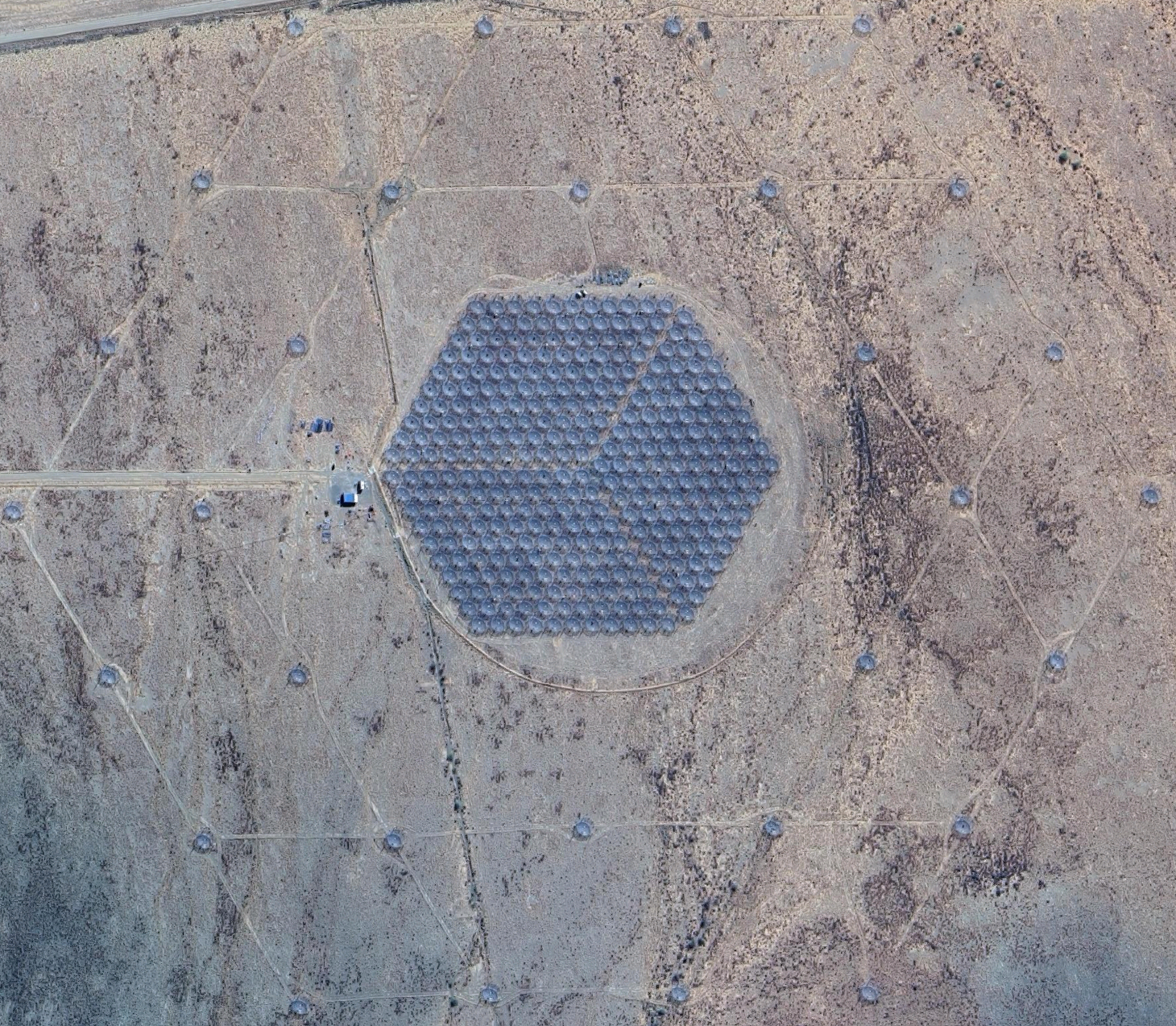}
    
    \caption{(top) The fully constructed hexagonal core. A Meerkat dish can be seen in the background. (bottom) The HERA array from an overhead Google Maps view. The 3 sectors of the hexagonal array and 2 layers of outriggers can be seen.}
    \label{fig:core}

\end{figure*}

Simulations suggest a wide range of possible models for the formation of the first stars and how emerging sources of radiation might heat and ionize the IGM. Observations of the redshifted neutral hydrogen (HI) emission or absorption signature trace the neutral gas regions through cosmic dawn and reionization, offering a window into the early universe (\citealp{Furlanetto_2006}, \citealp{Morales_2010}, \citealp{Pritchard_2010}). 

\begin{figure*}[!ht]
    \centering
    \includegraphics[width=1.8\columnwidth]{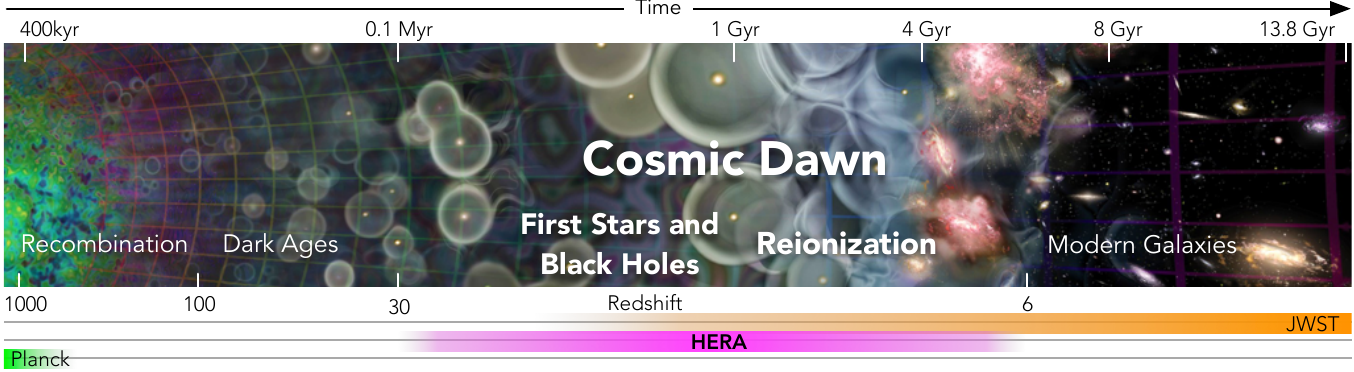}
    \caption{Cosmological timeline from just after the big bang to present day, highlighting the epochs of interest for HERA.}
    \label{fig:timeline}
\end{figure*}

Other experiments searching for the high redshift 21 cm signal include the Precision Array for Probing the EoR (PAPER, \citealp{Parsons_2010}), the Giant Metrewave Radio Telescope (GMRT, \citealp{gmrt}), the Murchison Widefield Array (MWA, \citealp{Tingay_2013} \& \citealp{Wayth_2018}), the LOw Frequency ARray (LOFAR, \citealp{2013A&A...556A...2V}), the Canadian Hydrogen Intensity Mapping Experiment (CHIME, \citealp{Newburgh_2014}), and the Large-Aperture Experiment to Detect the Dark Age (LEDA, \citealp{Price_2018}). Upcoming experiments include the Square Kilometer Array (SKA, \citealp{Mellema_2013}), The Canadian Hydrogen Observatory and Radio-transient Detector (CHORD, \citealp{2019clrp.2020...28V}), and the Hydrogen Intensity and Real-time Analysis eXperiment (HIRAX, \citealp{https://doi.org/10.48550/arxiv.2101.06338}).

HERA is an interferometer consisting of 350 parabolic 14-meter dishes, organized into a 320 antenna core and a 30 antenna outrigger configuration. The constructed core is shown in figure \ref{fig:core}, from a side view (top) and an overhead view (bottom). The 2 layers of outrigger antennas are also visible in the overhead view. This design is optimized for power spectrum detection as the redundant configuration bolsters sensitivity on the baselines corresponding to reionization and cosmic dawn, and the dishes provide a large collecting area. The first stage (``phase I'') of HERA reused the dipole feeds, cables, Post Amplifier Modules (PAMs), ROACH2 signal processing boards with 2x16 input ADCs, and an xGPU correlator from its predecessor, the Precision Array for Probing the Epoch of Reionization (PAPER), and refit the Front End Modules (FEMs) to be a 75 Ohm version of the future HERA FEM in order to match the impedance of the PAPER system. The phase I array consisted of a limited number of antennas. The ``phase II'' experiment upgrades the feed to a wideband design and replaces the digital and analog signal chains entirely. Additionally, phase II will encompass the full planned 350 antennas. This paper will describe the design and operation of the phase II array.

\section{Science with HERA-350}
\subsection{Primary Science Justification: From Cosmic Dawn to Reionization} \label{sec:sciback}

HERA's primary scientific goal is to understand the processes driving the evolution of the 21\,cm brightness temperature of the IGM during cosmic dawn and reionization. Figure \ref{fig:timeline} highlights the epochs of interest on a cosmological timeline. The array observes redshifts associated with cosmic dawn and reionization by tracing 21\,cm emission from neutral hydrogen. The 21\,cm hyperfine transition line is produced when a neutral hydrogen atom undergoes a spin-flip transition and emits a photon at a characteristic wavelength of 21\,cm. As the first astronomical objects formed, they emitted radiation that ionized the primordial intergalactic medium (IGM). By observing the fluctuations of the 21\,cm emission over time, HERA provides a unique tool to study the processes governing the early universe. HERA-350 has been optimized for detecting and characterizing the three dimensional power spectrum of the cosmological 21\,cm signal, where two dimensions map transverse to the line-of-sight and one maps redshift to line-of-sight distance. 

\begin{figure}[!ht]
    \centering
    \includegraphics[width=\columnwidth]{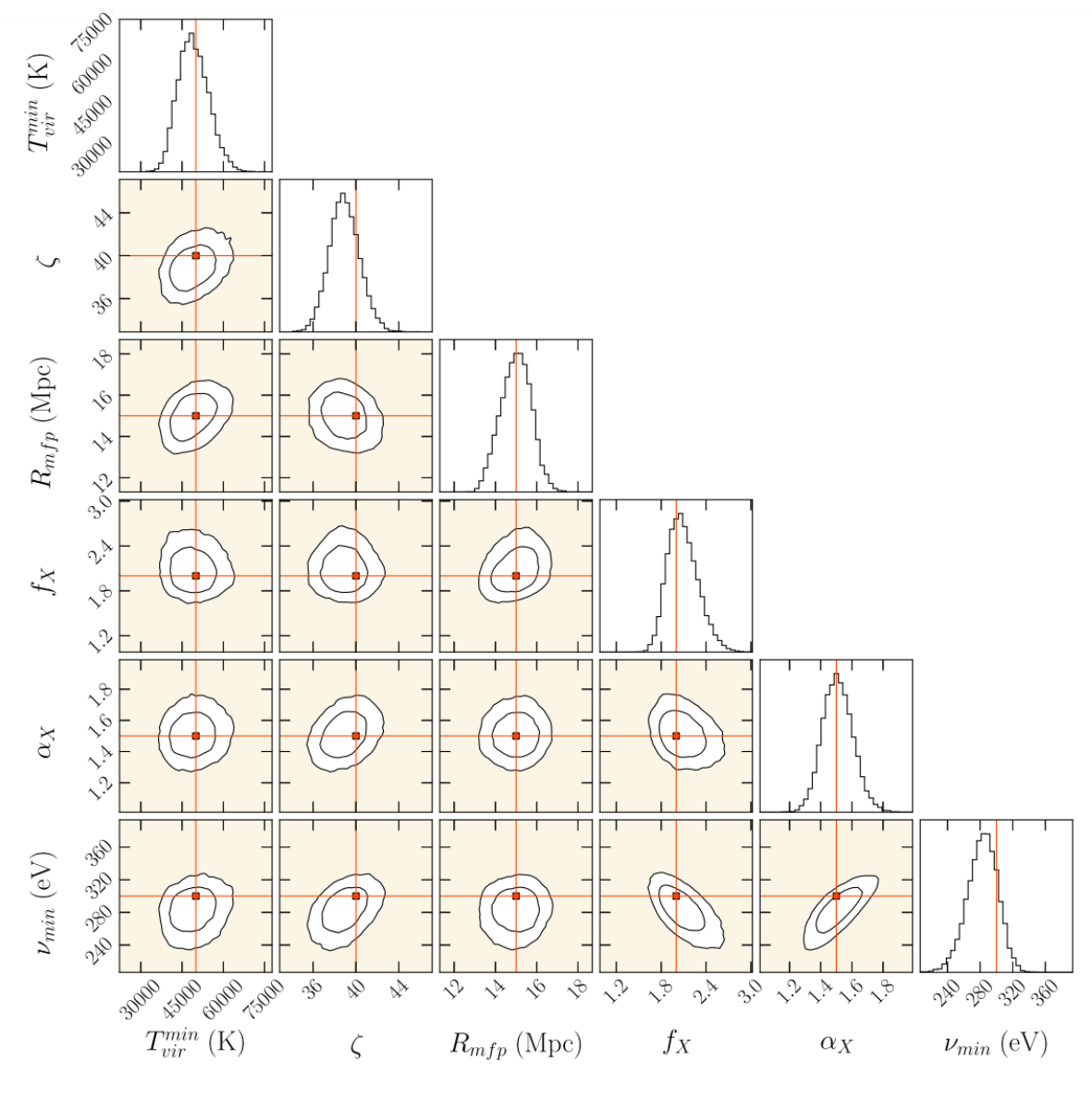}
    \caption{68\% and 95\% credible intervals for six key Cosmic Dawn parameters. The marginalized distribution across each model parameter is shown on the diagonal. EoR parameters include: the ionizing efficiency of star-forming galaxies ($\zeta$); mean free path of ionizing photons in the IGM ($R_{\rm mfp}$); and minimum virial temperature (i.e. mass) of galaxies contributing to reionization ($T_{\rm vir}^{\rm min}$).  Parameters covering heating during Cosmic Dawn include: X-ray production efficiency for early star formation ($f_{\rm X}$); spectral slope/hardness of the first X-ray sources $(\alpha_{\rm X}$); and minimum frequency of X-rays not absorbed by the ISM ($\nu_{\rm min}$). The crosshairs mark the true parameter of the mock observation used in the forecast. Existing constraints on these parameters are weak \citep{greig_mesinger2017}; however, HERA will nominally be able to constrain them to within $\sim$10\% with foreground-avoidance techniques. This figure is reproduced from \citealp{kern_et_al2017} with permission from the author.}
    \label{fig:MCMC}
\end{figure}

Figure~\ref{fig:MCMC} shows the results of a Markov chain Monte Carlo (MCMC) pipeline for fitting models to emulated multi-redshift 21\,cm power spectrum data, reproduced from \citealp{kern_et_al2017}.  The underlying models are based on the excursion-set formalism of \citealp{furlanetto_et_al2004} and the 21\textsc{cmfast} code \citep{mesinger_et_al2011}.  HERA-350 delivers $\lesssim$10\%-level constraints on these parameters, which are essentially unconstrained by current observations \citep{greig_mesinger2017}, especially those describing the pre-reionization heating epoch (X-ray production efficiency for early star formation ($f_{\rm X}$) and spectral slope/hardness of the first X-ray sources $(\alpha_{\rm X}$)). For more detail about the forecast, see \citealp{kern_et_al2017}. Other HERA forecasts, for example, include \citealp{Mason_2023} and \citealp{greig2017forecast}. 

\begin{figure*}[!ht]
    \centering
    \includegraphics[width=1.8\columnwidth]{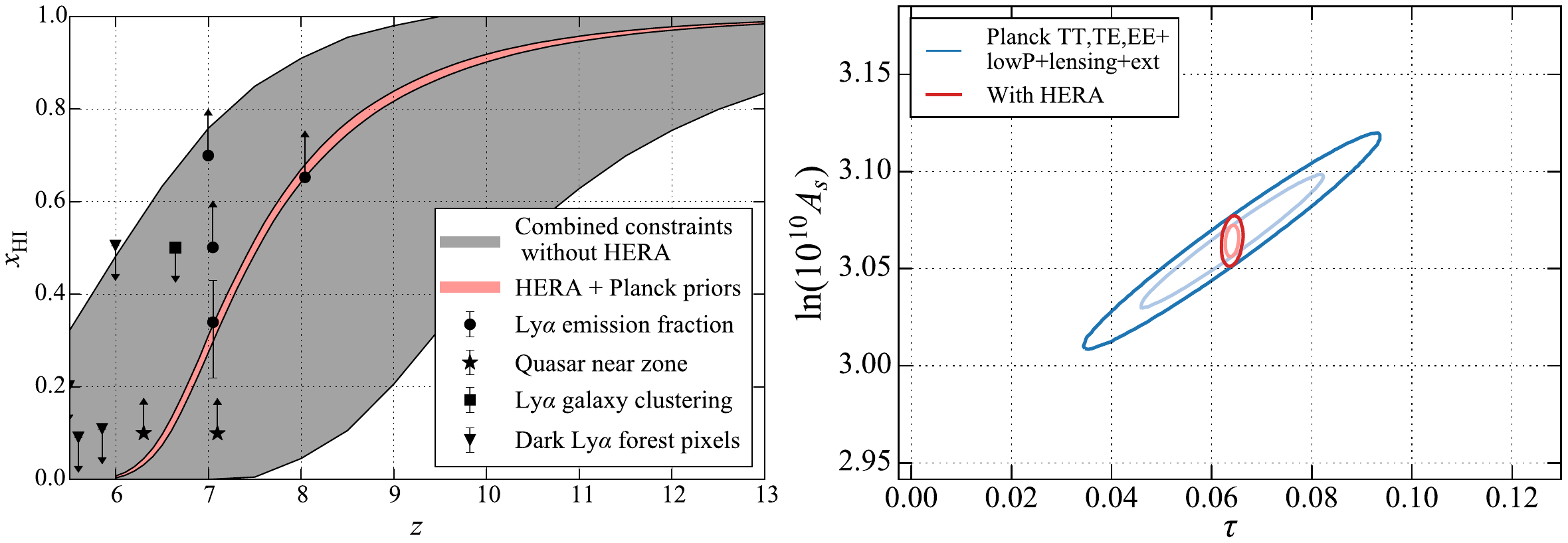}
    \caption{(left) Ionization history constraints from current high-$z$ observational probes (black points). With \emph{Planck} priors \citep{greig_mesinger2017}, the inferred $95\%$ confidence region (gray) reduces to the red region by adding HERA-350 measurements. (right) HERA-350's ionization history constraints break CMB parameter degeneracies, enabling improved constraints on $A_s$, $\sigma_8$, and the sum of the neutrino masses. For more details about these parameter forecasts for HERA, see \citealp{2016Liu}.}
    \label{fig:IonHist}
\end{figure*}

The telescope has had two phases of operation which differed in practical terms in the redshift range measured: phase I of HERA covered 6$<$z$<$13 and phase II extends this range to 5$<$z$<$27, encompassing the redshifts expected of the cosmic dawn. The HERA collaboration's first upper limits on the power spectrum of 21 cm fluctuations at a redshift of approximately 8 and 10 from phase I data have already been published (\citealp{phaseIlimits}, \citealp{heraresults2}), and the most sensitive limits are mostly consistent with thermal noise over a wide range of \textbf{k} (wavenumber).

While HERA's results are the most sensitive in the field at the time of this paper, they do not yet report a detection of the EoR. However, these results were extracted from data with a small subset of the array, only 39 of 52 working antennas, and the observations used the phase I signal chain described in \citealp{DeBoer_2017}. Future HERA results will be published with data from the phase II signal chain described in this paper, and will use more antennas.

\subsection{Secondary Science Goal: Constraints on Fundamental Physics}
HERA's constraints on reionization history also help 
constrain fundamental physics. As a
direct probe of reionization, HERA observations can remove the optical depth
$\tau$ as a nuisance parameter in CMB studies. Though the inference of $\tau$ from the 21 cm data is model-dependent, 
such a measurement can still have a significant impact.  As shown in figure \ref{fig:IonHist}, 
knowing $\tau$ breaks 
the internal CMB degeneracy with
the amplitude of primordial scalar fluctuations $A_s$,
effectively reducing errors on $A_s$ by a factor of four
\citep{liu_et_al2016}. For more details about forecasts for HERA, see \citealp{2016Liu}. 

Including HERA
constraints on $\tau$ also reduces the errors on $\sum m_\nu$, the sum of the neutrino masses, to $\sim$12 meV,
providing a $\sim$5$\sigma$ detection of the neutrino mass even 
if $\sum m_\nu$ $\sim 58\,\textrm{meV}$, the current minimum allowed value.
The breaking of degeneracies provided by $21\,\textrm{cm}$ measurements become
even more useful for measuring $\sum m_\nu$ if future cosmological datasets
demand more than the current six parameters of $\Lambda$CDM. 

\subsection{Secondary Science Goal: Cross Correlations with other Early Universe Probes}

In addition to statistical power spectra,  deep, foreground-cleaned image cubes, spanning
$0.8\times 0.8\times 18$~Gpc$^3$ and $5< z < 27$ with
$\Delta z/z<0.05$, could provide additional information when combined with current and future datasets.  In principle, informative cross-correlations using higher-point statistics are possible with patchy reionization from the kinetic SZ effect \citep{McQuinn_2005}. Galaxy populations, characterized by spectroscopic tracers and the intensity mapping of lines such as Ly$\alpha$, Ly-break, CO, or C\textsc{ii}, provide another cross-correlation opportunity (\citealp{mcbride}, \citealp{beane}). 

On large scales, the cross-power spectra between these emission lines and the  $21\,\textrm{cm}$ line are negative because the cross-correlations are driven by fluctuations in ionization. On small scales, all lines trace density fluctuations, giving rise to positive cross-correlations \citep{gong_et_al2012}. Measuring the scale at which the correlations change sign provides robust information on both the ionization state of the IGM and the characteristic spatial scale of ionized bubbles. The former provides independent confirmation of the ionization history constraints provided by HERA alone, while the latter allows one to \emph{test}---rather than assume---theoretical models of the effects of inhomogeneous recombination in the ionized IGM. Moreover, cross-correlations provide an invaluable opportunity to test one's data for residual foreground signals and other systematic errors.

Though not optimized for imaging, the HERA team is investigating foreground subtraction and imaging techniques, which will produce datasets that could be used in combination with other high redshift tracers. Foreground subtracted images can be compared with or cross-correlated against other high redshift observables such as the Ly$\alpha$ forest (e.g with SPHEREx, \citealp{Cox_2022}), CO and CII lines (e.g, \citealp{Gong_2011}, \citealp{Carilli_2011}), the kinetic Sunyaev-Zeldovich effect (e.g, with the Simons Observatory, \citealp{La_Plante_2020}), and even direct observations of high redshift galaxies (e.g, with existing surveys, \citealp{pagano2021}, with JWST, \citealp{Beardsley_JWST}, or with the Nancy Grace Roman Space Telescope, \citealp{La_Plante_2023}).

\begin{figure}[ht]
    \centering
    \includegraphics[width=\columnwidth]{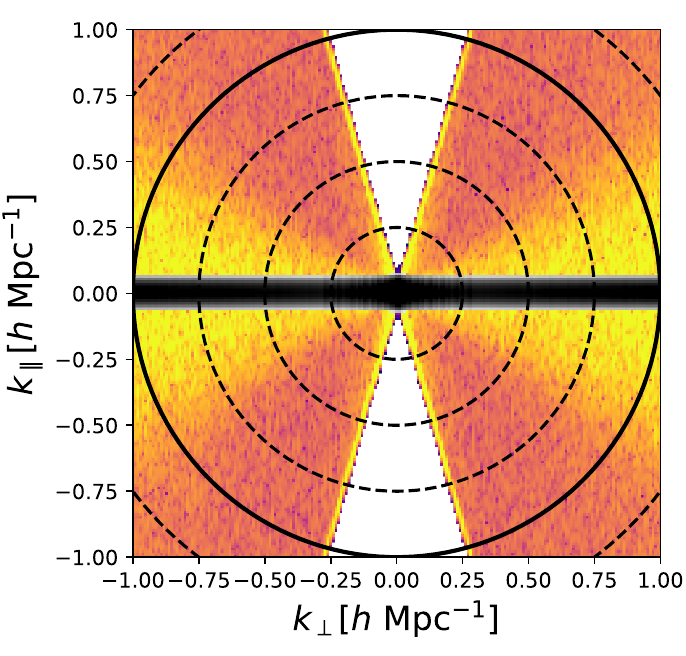}
    \caption{Smooth-spectrum foregrounds up to $10^6$ times brighter than the EoR signal are intrinsically confined to low-order $k_\parallel$ modes (black region). The chromatic response of an interferometer scatters power in a characteristic wedge shape (colored region). HERA's foreground avoidance strategy uses smooth instrumental responses to produce a foreground-free window (white) for detecting the EoR.}
    \label{fig:wedge}
\end{figure}

\section{Approach to Power Spectrum Measurements} \label{sec:pspec}

The HERA-350 design was optimized to maximize sensitivity to the power spectrum while minimizing the impacts of bright foregrounds. 

One of the difficulties faced by EoR experiments is that of strong astronomical foregrounds, which are expected to be up to 6 orders of magnitude higher than the EoR signal \citep{Santos_2005}. These foregrounds are mainly synchotron and free-free radiation from our own galaxy and other galaxies, which are spectrally smooth phenomena.  HERA takes a delay spectrum approach to statistical power spectrum calculation, where the power spectrum is first calculated as the Fourier transform along frequency for each baseline before combining data from many baselines. Smooth spectrum emission is isolated to a small number of Fourier modes, while the sharp-edged 21 cm background persists to higher delay modes. HERA-350, with redundant baselines sensitive to Fourier modes of interest, is designed to take advantage of this analysis method. However, other power spectrum techniques are under investigation as well and will be discussed briefly. 

Though 21 cm observations offer a full 3D probe of the early universe, a 2-D cylindrical average is a useful space to discuss how array design approaches impact sensitivity in Fourier space. Figure \ref{fig:wedge} illustrates how foregrounds and background map to line-of-sight $k_\parallel$ modes, and perpendicular to line-of-sight modes $k_\perp$. The range of $k_\parallel$ is set by the instrument bandwidth as the lower bound and the spectral resolution as the higher bound, while each baseline type samples a different $k_\perp$ mode. The longest baselines probe the finest angular scales at the upper bound, and the lower bound is determined by the shortest baselines. As HERA is a compact array with dishes touching nearly edge to edge, this bound is equivalent to the 14.6 meter dish spacing. 

Foreground avoidance takes advantage of the expected characteristics of the foregrounds and the EoR signal in k-space. Foregrounds are expected to be spectrally smooth and are therefore confined to the lower order $k_\parallel$ modes. However, this natural sequestering of foregrounds in k-space is disturbed by instrumental effects. Interferometer baselines probe specific angular scales scaling inversely with baseline length; however, these scales also vary with frequency. Because of this inherent chromaticity, foregrounds spread beyond the lowest $k_\parallel$ modes, forming a wedge shape in 2D k-space (see figure \ref{fig:wedge}). The extent of this wedge may also be impacted by systematics, such as calibration errors, which can spread foreground power outside of the wedge and into the EoR window \citep{Barry_2019}.  The wedge and other mode mixing effects are theoretically discussed in  \citealp{Morales_2012}, \citealp{vendatham_2012}, \citealp{Parsons_2012_2}, \citealp{Hazelton_2013}, \citealp{Thyagarajan_2013}, \citealp{dillon_2013},
\citealp{2014liu},
\citealp{2014liu2},
\citealp{Thyagarajan_2015},
and \citealp{2016Liu2}. The leading approach in most analyses is to minimize instrumental wedge-causing systematics and then avoid including any areas which do become contaminated. 

\begin{figure*}[ht]
    \centering
    \includegraphics[width=2\columnwidth]{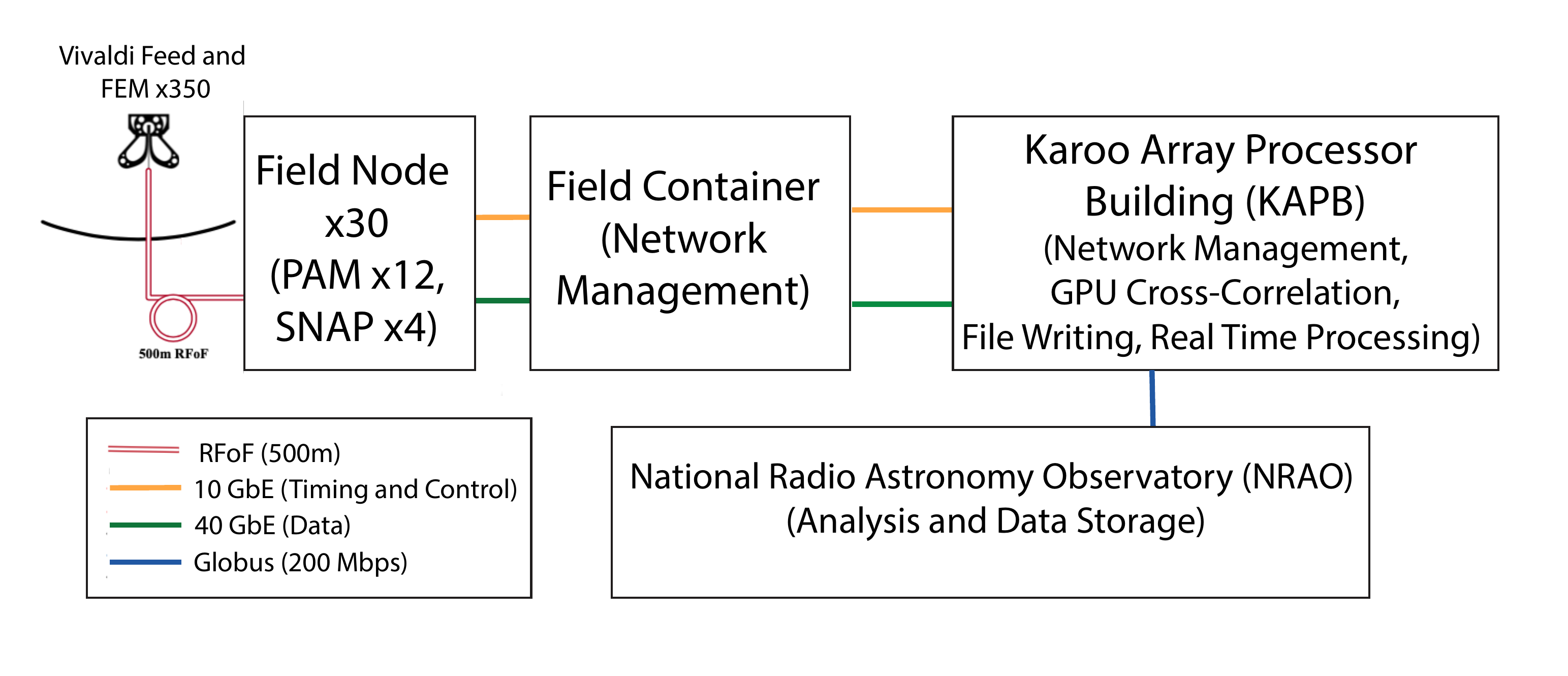}
    \caption{HERA’s signal path from antenna to data archive. Front end modules (FEMs) embedded in the feed convert RF signals to optical fiber and send them to the Post Amplifier Modules (PAMs) over a 500-meter RF over Fiber (RFoF) connection. Signals are received in one of 30 nodes distributed about the array. Smart Network ADC Processors (SNAPs) digitize, Fourier transform, and output packets over optical fiber, which are sent over a 10 km fiber bundle to the Karoo Array Processing Building (KAPB). The KAPB hosts the GPU correlator, which cross multiplies, averages to 100ms, and sends products to a data catcher for further averaging and writing to disk. The Real-Time Processor (RTP) calibrates and flags data on a nightly cadence. Final products are transferred to the National Radio Astronomy Observatory (NRAO).}
    \label{fig:systemoverview}
\end{figure*}

Figure \ref{fig:wedge} demonstrates the characteristic features in k-space. The foregrounds (black) are intrinsically confined to low $k_\parallel$ modes, however, instrument chromaticity spreads their power into other modes, creating a characteristic wedge shape (color region). The white area represents the foreground free EoR window.  Contours indicate 21 cm power spectrum amplitude (P(k)) which is predicted to peak at the smallest k-modes.

The HERA approach is to discard power spectrum bins containing foregrounds (any colorized pixel in Fig \ref{fig:wedge}). The array design focuses on maximizing sensitivity on the non-foreground contaminated modes with a large collecting area and a highly-compact redundant configuration. The redundancy offers a particularly large sensitivity boost \citep{parsons_2012}. Additionally, noise averages down faster when coherently combining visibilities than it does when combining power spectrum measurements. This approach and a consideration of the tradeoffs is discussed in \citealp{Dillon_2016}. 

The delay spectrum technique is the primary approach for HERA analysis pipelines. It has the advantage of simplicity, since it turns visibilities (the natural measurement basis of an interferometer) directly into power spectra without intermediate steps. However, there are a number of benefits to pursuing imaging based pipelines. Images are interesting in their own right for science, as well as for testing array performance and mapping foregrounds. Rotation and spectral synthesis mean the same point in UV space is measured multiple times, even with different baselines. This improves sensitivity over the delay spectrum approach as we are able to combine information from non-identical baselines. These types of analyses do not require raw visibilities to be stored after map-making, meaning the data is significantly compressed at this stage, which can help reduce data storage requirements. Imaging can also assist with recovery of Fourier modes contaminated by foregrounds. In order to obtain measurements in the contaminated regions of k-space, foreground subtraction is necessary. While a foreground model can be simulated and subtracted as a part of the delay spectrum analysis, it is more naturally part of an imaging analysis where errors in source model amplitude can be tracked to locations on the sky. 

For a more in depth discussion of the techniques and tradeoffs of various EoR analyses, see \citealp{Morales_2018} and \citealp{Liu_2020}. 

\begin{figure*}[ht]
\centering
\includegraphics[width=2\columnwidth]{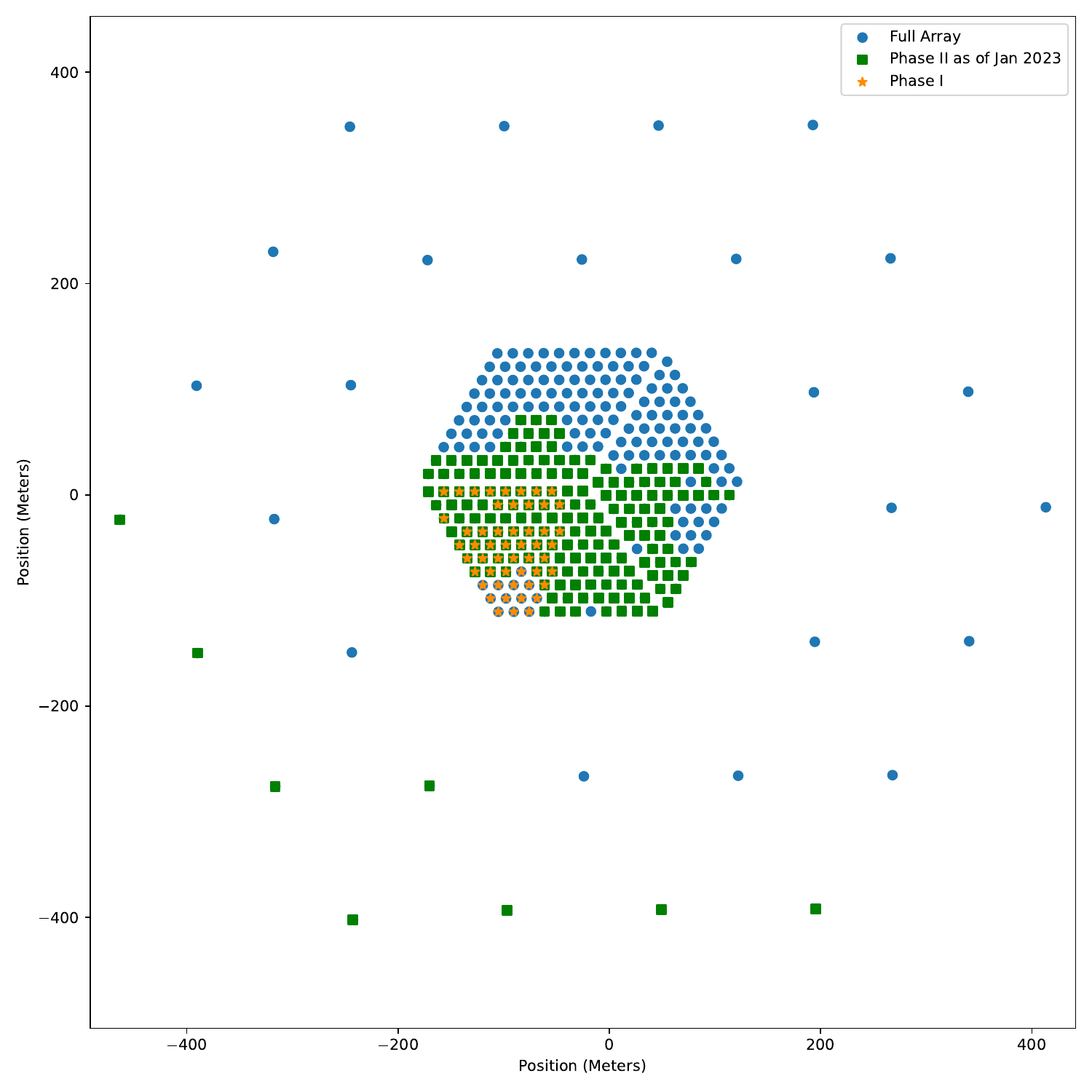}
\caption{The HERA array layout at a few stages of construction. The phase I sub-array, and the phase II array as of Jan 2023 are highlighted. The full array configuration is underplotted.}
\label{fig:layout}
\end{figure*}

\section{The HERA Telescope} \label{sec:sysdesign}
HERA is located at the South African Radio Astronomy Observatory site outside the town of Carnarvon in the Northern Cape of South Africa. This is also the site of the Meerkat telescope and the future site of the SKA Mid array. HERA was built in two phases. The first phase was built with parts recycled from the PAPER experiment previously operated at the same location. In this configuration the telescope was limited to redshifts 6$<$z$<$13 and could support at most 75 antennas. In the second phase a feed and signal chain upgrade extended the redshift limits to 5$<$z$<$27 and the digital system was upgraded to support all 350 antennas. Near real time analysis steps such as flagging and calibration have been improved to better support rapid detection and diagnostics of system issues and have led to faster turn-around of subsequent data analysis.

\begin{figure*}
\centering
\includegraphics[width=2\columnwidth]{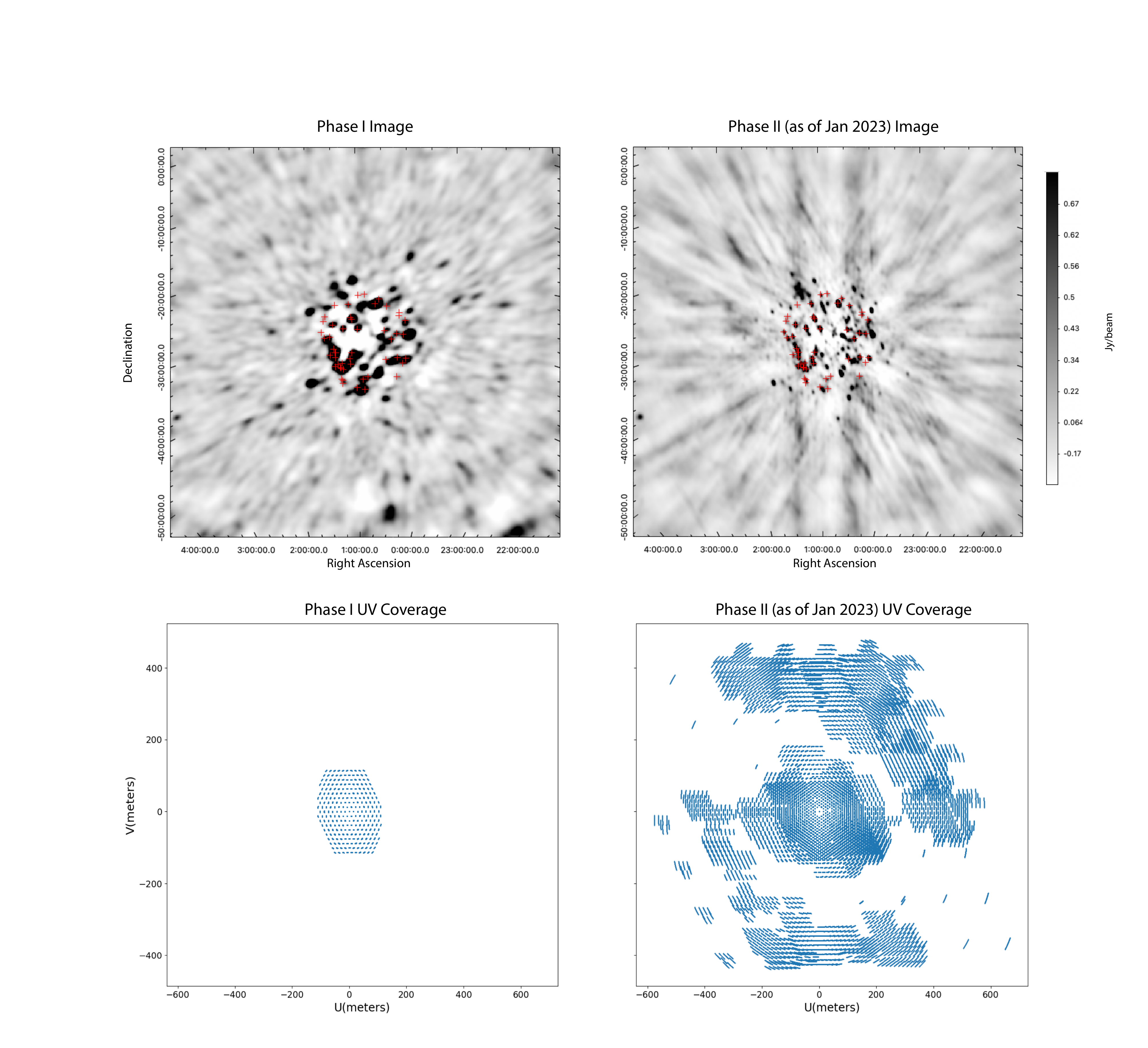}
\caption{(top) Synthesis images demonstrating the HERA configuration and overall performance. (top, left) Phase I with 52 antennas and (top, right) Phase II with 183 antennas, as of January 2023. The red crosses indicate the top 50 brightest sources in peak flux at 122 MHz in the GLEAM catalogue (\cite{gleam}) within 7 degrees of the image center. Sources that appear bright in the image but not in the top brightest 122 MHz GLEAM catalog sources are likely blended sources. Roughly 20 minutes of data and a frequency range of 100-200 MHz was used to make the images. The phase I image shows clearly less well localized point sources, as well as less sources in total. The H6C image shows improved resolution. (bottom) Corresponding UV coverage for the images in the top panel. (bottom, left) Phase I with 52 antennas and (bottom, Right) Phase II with 183 antennas. The H6C image covers a broader UV range, and contains a denser UV plane.}
\label{fig:uvcomp}
\end{figure*}

An overview of the full system is included in figure \ref{fig:systemoverview}, showing the array support structures and their roles. Each of the 350 array elements is a 14-meter mesh dish with a suspended feed that contains an embedded amplifier module. This Front End Module (FEM) acts as balun, amplifier and RF over fiber converter. The main RF signal is sent via 500m of fiber to one of 30 nodes where it is converted back to radio frequency by a Post Amplifier Module (PAM) and then digitized by a Smart Network ADC Processor (SNAP). The 500-meter length was chosen in order to place reflections well outside of delays corresponding to cosmological modes of interest. The SNAP boards output packets of channelized spectra which are routed to the GPU correlators for cross-correlation in the nearby Karoo Array Processor Building (KAPB), which also hosts the cluster for real-time processing (RTP) of first round flagging and calibration. Once the RTP has finished and data products are written, they are transferred off site to the National Radio Astronomy Observatory in Socorro, NM for analysis and long term storage.

\subsection{Array Layout} \label{sec:layout}

The HERA array layout is shown in figure \ref{fig:layout}, with the phase I sub-array and the phase II array as of Jan 2023 highlighted. The array is arranged using hex packing which allows neighboring dishes to share supports. The stationary, zenith pointing, dishes are arranged in a compact hexagonal core, with dishes nearly touching end to end and 14.6 meter center-to-center spacing. This densely-packed redundant layout maximizes sensitivity on a small number of modes, especially short baselines---as desired based on the approaches laid out in section \ref{sec:pspec}---though this also makes the array less suited to tomographic mapping. While HERA primarily targets delay spectrum based approaches, investigations into alternative imaging based power spectrum pipelines are under development. Therefore, a few design choices were made to improve imaging capabilities. This motivated the addition of two rings of outriggers giving longer baselines, and motivated splitting the core 320 antennas into three sub-sections. Each sub section is moved with respect to the others by 1/3 of a grid spacing to fill in the repeating gaps in the $uv$ plane.  

Another benefit of regularly spaced antennas is the possibility of redundant calibration (\citealp{firstredcal}, \citealp{redcal21cm}). We can take advantage of the fact that grids of antennas make many measurements of the same sky fringe. This situation significantly reduces the number of free parameters to be constrained by a sky model.
Despite the split core configuration and outriggers, HERA's full array still can be calibrated redundantly, leaving only four free parameters per frequency and polarization (one parameter each for amplitude, overall phase, East-West tip/tilt, and North-South tip-tilt) which must be solved for with the introduction of the method described in \citealp{Dillon_2018}. In practice it has been found that sky model errors can still introduce unwanted chromaticity (\citealp{barry2016}, \citealp{Ewall_Wice_2017}, \citealp{byrne2019}) but this can be partially mitigated by down-weighting long baselines \citep{Ewall_Wice_2017, Orosz_2019}, gain smoothing \citep{Kern_2020b}, or an approach which combines redundant and non-redundant methods \citep{byrne2021}. HERA's outriggers and split-core configuration also makes it particularly suitable to calibration using spectral redundancy, which can dramatically reduce the number of free parameters after redundant baseline calibration from a few per frequency to just a handful over the whole band \citep{cox2023spectral}. 

Figure \ref{fig:uvcomp} shows a comparison of the sky imaged with the phase I and phase II commissioning arrays, as well as the UV plane coverage for the images. It is important to note that the phase II data shown here is not the full expected array, and contains only the antennas built as of January 2023. The antennas used for the images are highlighted as the respective sub-arrays in figure \ref{fig:layout}. The phase II array data used the new signal chains and contained a subset of the array with 183 antennas, as compared to 52 antennas with the old signal chains in the phase I data. The red crosses indicate the top 50 brightest sources in peak flux at 122 MHz in the GLEAM catalogue \citep{gleam} within 7 degrees of the image center. Sources that appear bright in the image but not in the top 50 brightest sources at 122 MHz in the GLEAM catalog are likely blended sources. All top GLEAM sources have an obvious counterpart in the HERA images, although they may appear faint for a number of reasons (e.g, the HERA image is wideband and the source may have less flux at other wavelengths, the source is not well localized) The images cover the same field and frequency range (100-200 MHz), as well as both using roughly 20 minutes worth of data. The images are made with the CASA package \citep{casa}. The phase II array image shows markedly improved resolution, shown by the better localization of point sources as well as the increase in the number of resolved sources. This can be attributed to the much broader UV plane coverage in the phase II data, even with only a portion of the phase II antennas built out.

\begin{figure}
    \centering
    \includegraphics[width=\columnwidth]{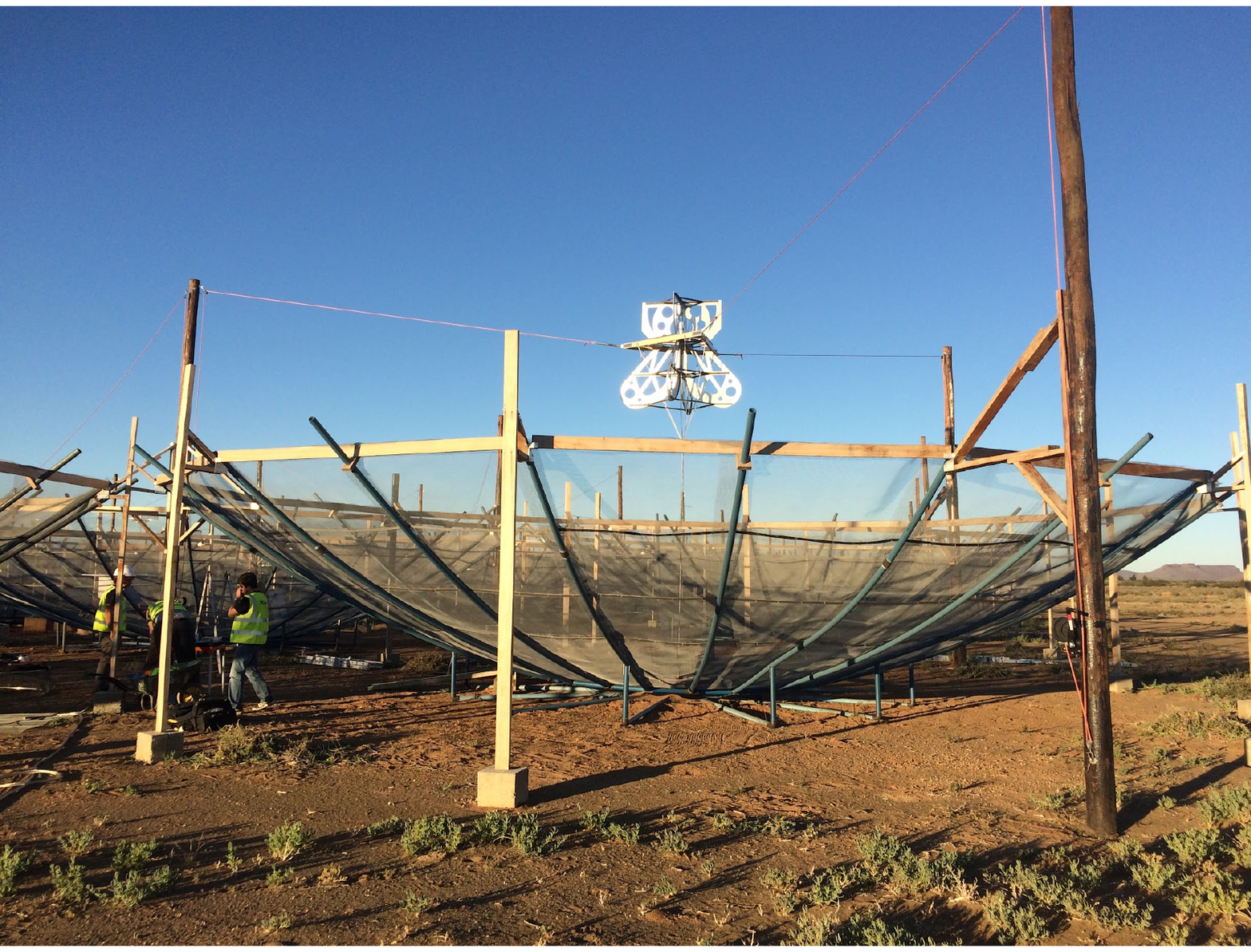}
    \caption{A single HERA dish. The Vivaldi feed is suspended by non conducting marine grade kevlar rope to poles on the perimeter of the dish. For servicing, the feed is lowered by hand winch and reached via a removable reflector panel.} 
    \label{fig:outrigger}
\end{figure}

\subsection{Dish and construction}
The HERA dish surface is a 5mm wire mesh optimized for reflecting wavelengths within HERA's bandwidth. It is supported by radial arms made from PVC pipe radiating from a concrete hub at dish center. The radials are supported at one point by vertical spars which are also tied back to the hub. Once weighted by mesh, the spars form a faceted parabola surface. One panel of the mesh contains a small removable door and bridge to allow for access into the dish. A detailed description of the dish construction and design can be found in \citealp{DeBoer_2017}.

\begin{figure}
    \centering
    \includegraphics[width=\columnwidth]{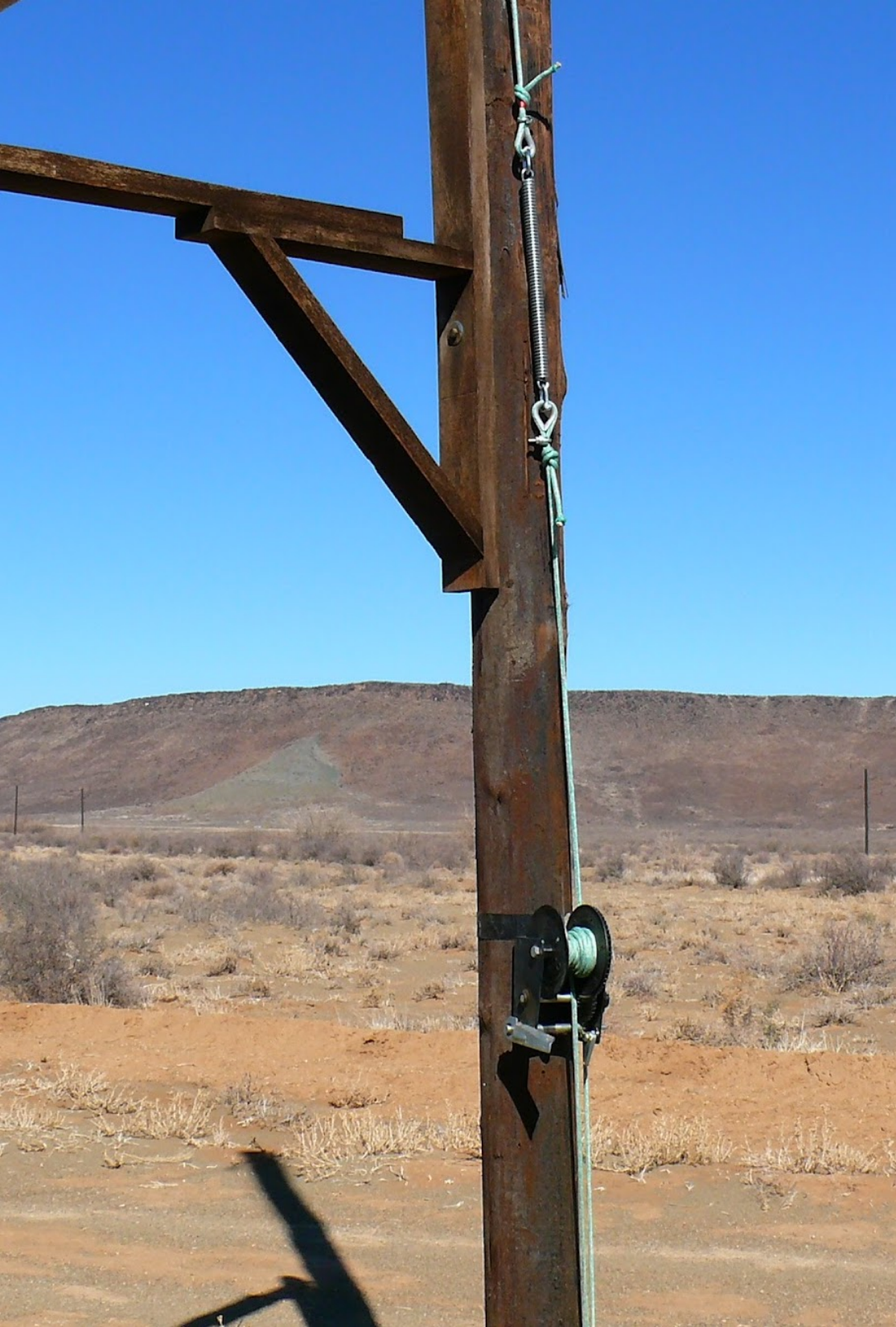}
    \caption{The feeds are raised and lowered with a hand winch. Each pole serves 3 surrounding antennas, therefore each pole also supports 3 winches, except for edge cases such as the one in this figure. Also visible is the extension spring which allows safe adjustment of tension force. Originally these were located at the feed attachment point, but were observed to introduce notable distortions in antenna response.}
    \label{fig:winch}
\end{figure}

The feed, described in the next section, is suspended from 3 Aramid fiber ropes strung via telephone poles. Three antennas share a pole to balance forces and reduce the total number of required supports. On the edges of the array perimeter poles are stayed with guy lines. The feeds are lofted via winches and high tension springs are inserted onto the ropes as a safety measure. Originally these springs were placed near the feed, but it was noted that this caused undesired resonances in the passband of the antenna, and they were moved near the poles.  An isolated view of an outrigger antenna is included in figure \ref{fig:outrigger} to more clearly show feed, rope, and pole construction. Each pole also has a winch per antenna to raise and lower of feeds for maintenance, and to allow trimming of feed heights across the array. This winch is shown in figure \ref{fig:winch}.

\subsection{Phase II Feed}
\begin{figure}
    \centering
    \includegraphics[width=\columnwidth]{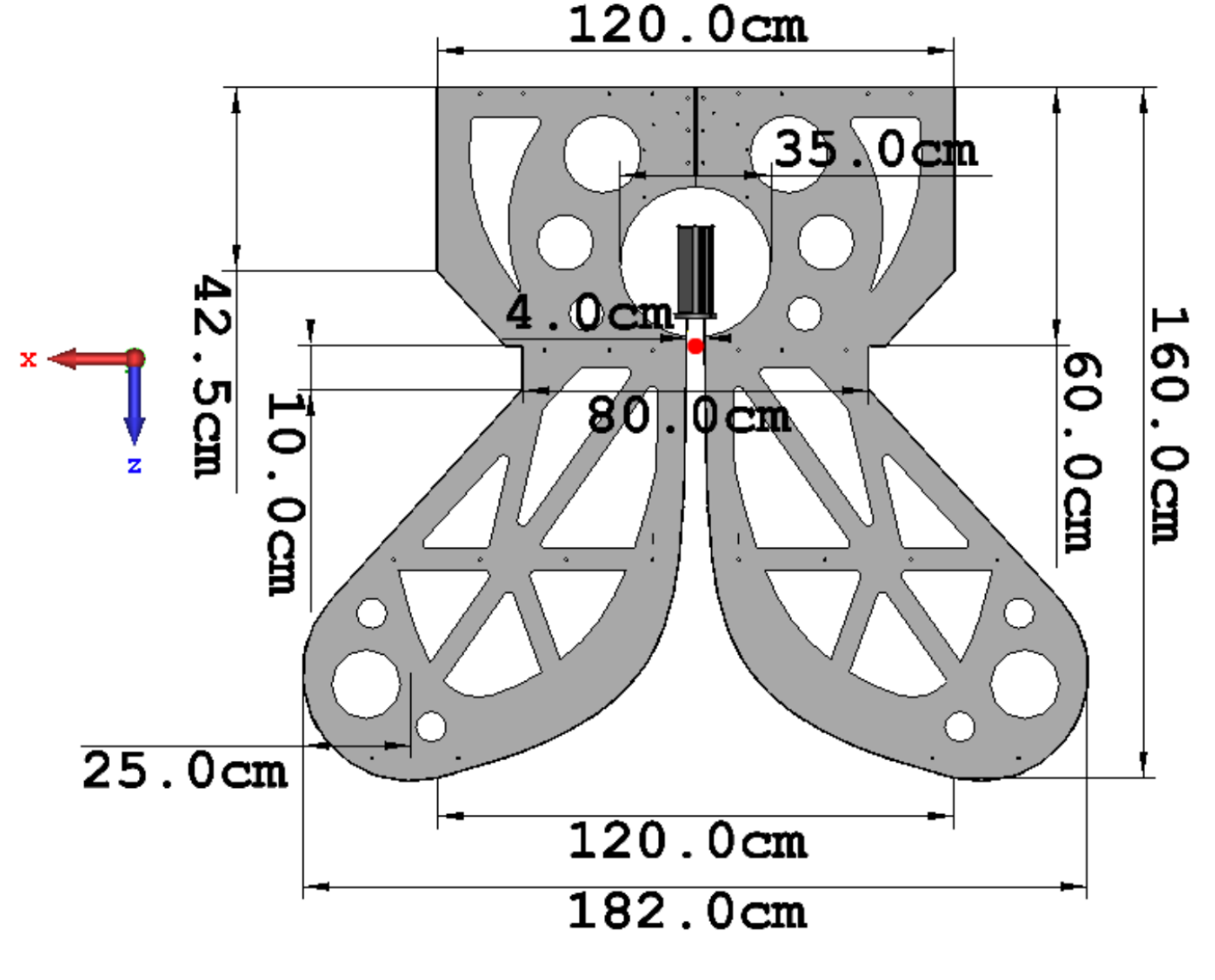}
    \caption{Dimensions of a Vivaldi blade, reproduced from \citealp{Fagnoni_2021} with permission from the author.}
    \label{fig:vivdims}
\end{figure}

The phase I iteration of HERA reused the dipoles from the PAPER experiment suspended over the 14-meter diameter zenith pointing dish. This feed had a frequency range of 100-200 MHz. Phase II replaced this feed with a Vivaldi type antenna which extended useful science bandpass down to include the Cosmic Dawn band at 50 to 100MHz (\citealp{Fagnoni_2021}, \citealp{de_Lera_Acedo_2020}), and the later stages of reionization above 200 MHz. The full range of the Vivaldi feed spans 50-250 MHz. The dimensions of the Vivaldi blade are shown in figure~\ref{fig:vivdims}.

\begin{figure}[ht]
    \centering
    \includegraphics[width=\columnwidth]{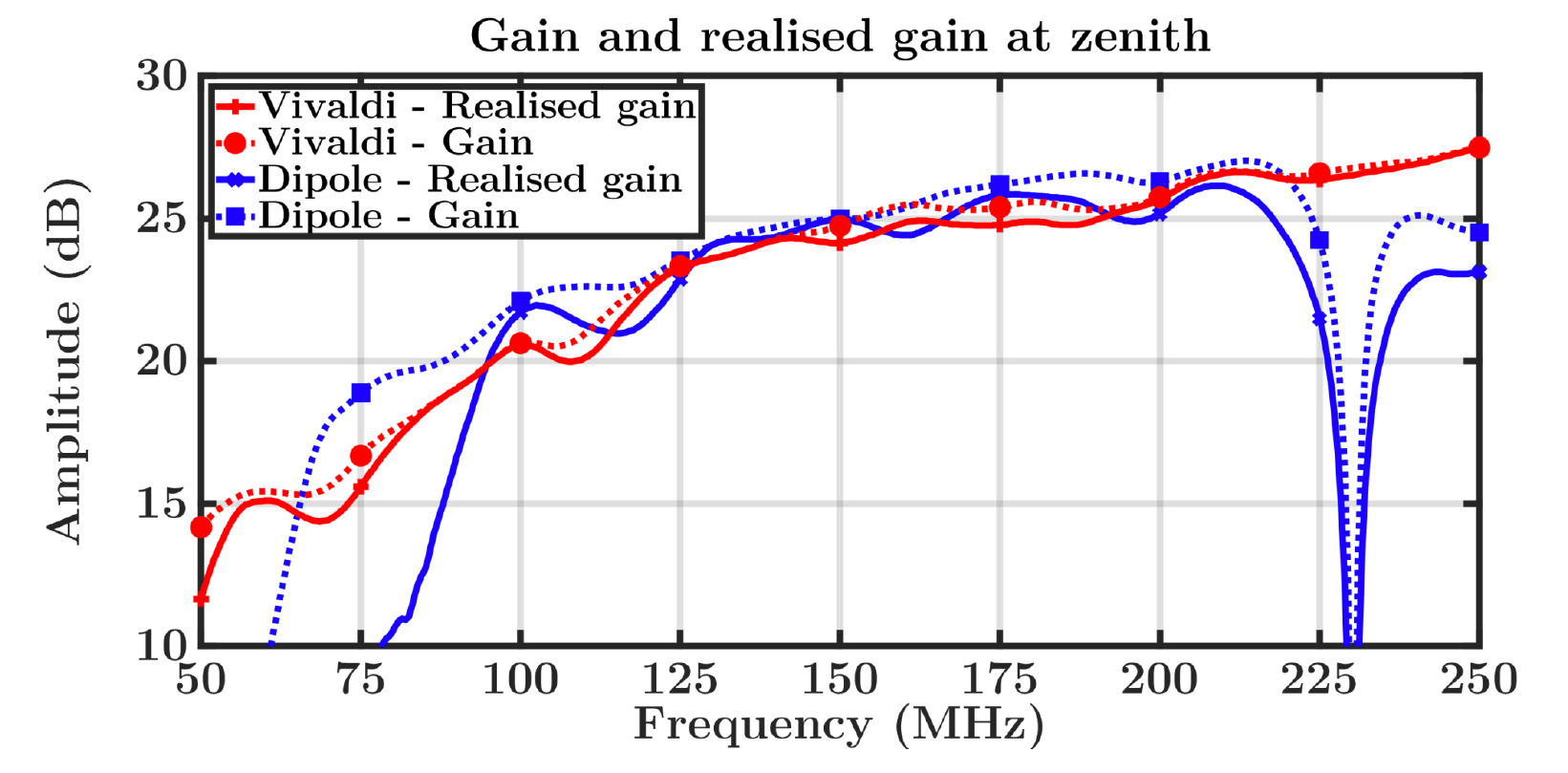}
    \caption{Gain and realized gain at zenith for the Vivaldi feed with upgraded Front End Modules (FEMs) indicated in red, as compared to the phase I HERA feeds (PAPER dipoles) and FEMs in blue. The extended bandwidth and spectral smoothness of the Vivaldi system is clearly seen. The gain on overlapping frequencies for the two systems is comparable. Reproduced from \citealp{Fagnoni_2021} with permission from the author.}
    \label{fig:vivresponse}
\end{figure}

\begin{figure}[ht]
    \centering
    \includegraphics[width=\columnwidth]{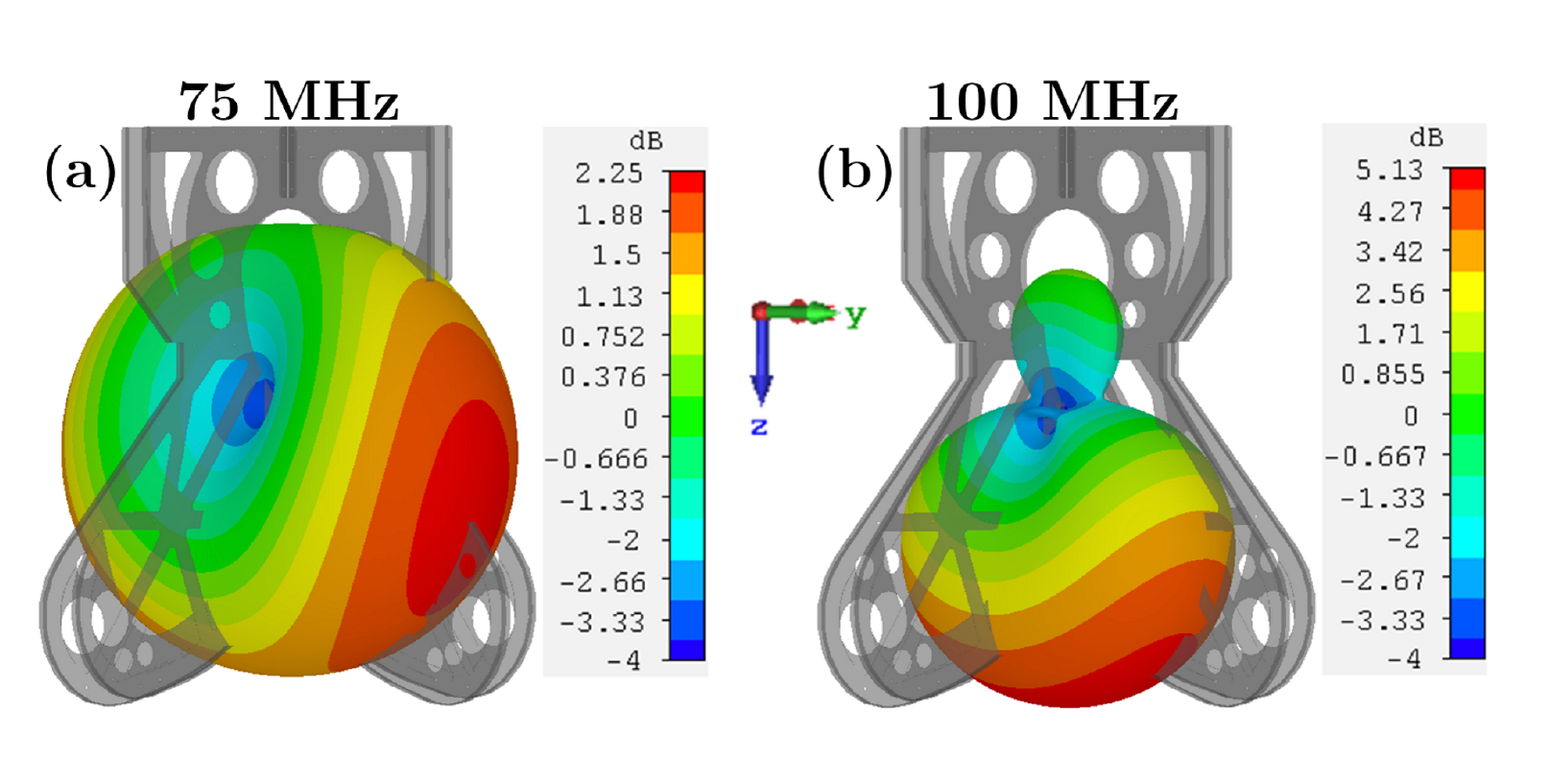}
    \caption{3-D beam pattern for the phase II Vivaldi feed X-polarization at 75 and 100 MHz. Reproduced from \citealp{Fagnoni_2021} with permission from the author.}
    \label{fig:beam}
\end{figure}

 The system uses a feed that does not require a backplane, which radiates towards the dish with minimal backwards gain. The shape and matching network are optimized to match between the feed output and the receiver. The Vivaldi feed and its position in the dish were designed to minimize unwanted spectral structure. For example reflections between dish and feed can cause ripples at frequency scales corresponding to target line-of-sight Fourier modes. The phase II feed has a lower reflection coefficient leading to lower standing waves within the dish. The gain of the antenna and the realized gain (combining the antenna directivity, radiation efficiency, and impedance mismatch with the receiver) are shown in figure~\ref{fig:vivresponse}. By removing the sharp cutoffs at 100 and 225 MHz the bandwidth has increased substantially while the spectral variation is kept to a similar level. The 3D beam pattern of the X-polarization of the feed at 75 MHz and 100 MHz is shown in figure \ref{fig:beam}. 

\begin{figure}
    \centering
    \includegraphics[width=\columnwidth]{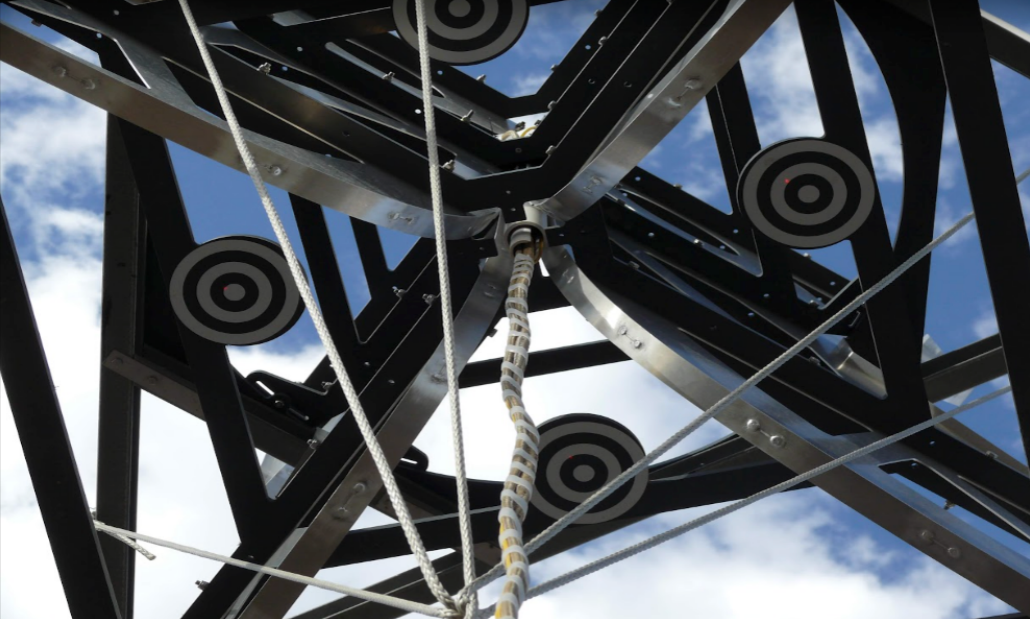}
    \caption{A view of the underside of a Vivaldi feed, showing the targets for the leveling jig. This jig ensures that the feed is positioned correctly using a laser plumb bob.  The cable wrapping coming from the feed center extends to the bottom of the image.}
    \label{fig:leveling}
\end{figure}

Numerical electromagnetic simulations showed the gain spectrum to be very sensitive to the positioning of the feed. This was subsequently confirmed experimentally by varying feed height and observing changes in autocorrelations. \citealp{Kim_2022} found in simulation that feed positioning should be accurate to the 1 centimeter level in position and 1 degree in tilt to minimize foreground leakage in power spectra. This leakage is caused by calibration errors, as the feed positioning offset impacts array redundancy. \citealp{Kim_2023} showed that with mitigation techniques, this requirement could be relaxed to 2 centimeters in position and 2 degrees in tilt. To control the feed position, a levelling jig with a laser plumb bob was devised. When placed on the dish hub, the levelling and distance lasers point at a set of targets on the feed seen in figure \ref{fig:leveling}.  With this system the feed can be positioned to better than 10mm.

\subsection{Analog Signal Path and Node} \label{analog}
\begin{figure}
    \centering
    \includegraphics[width=\columnwidth]{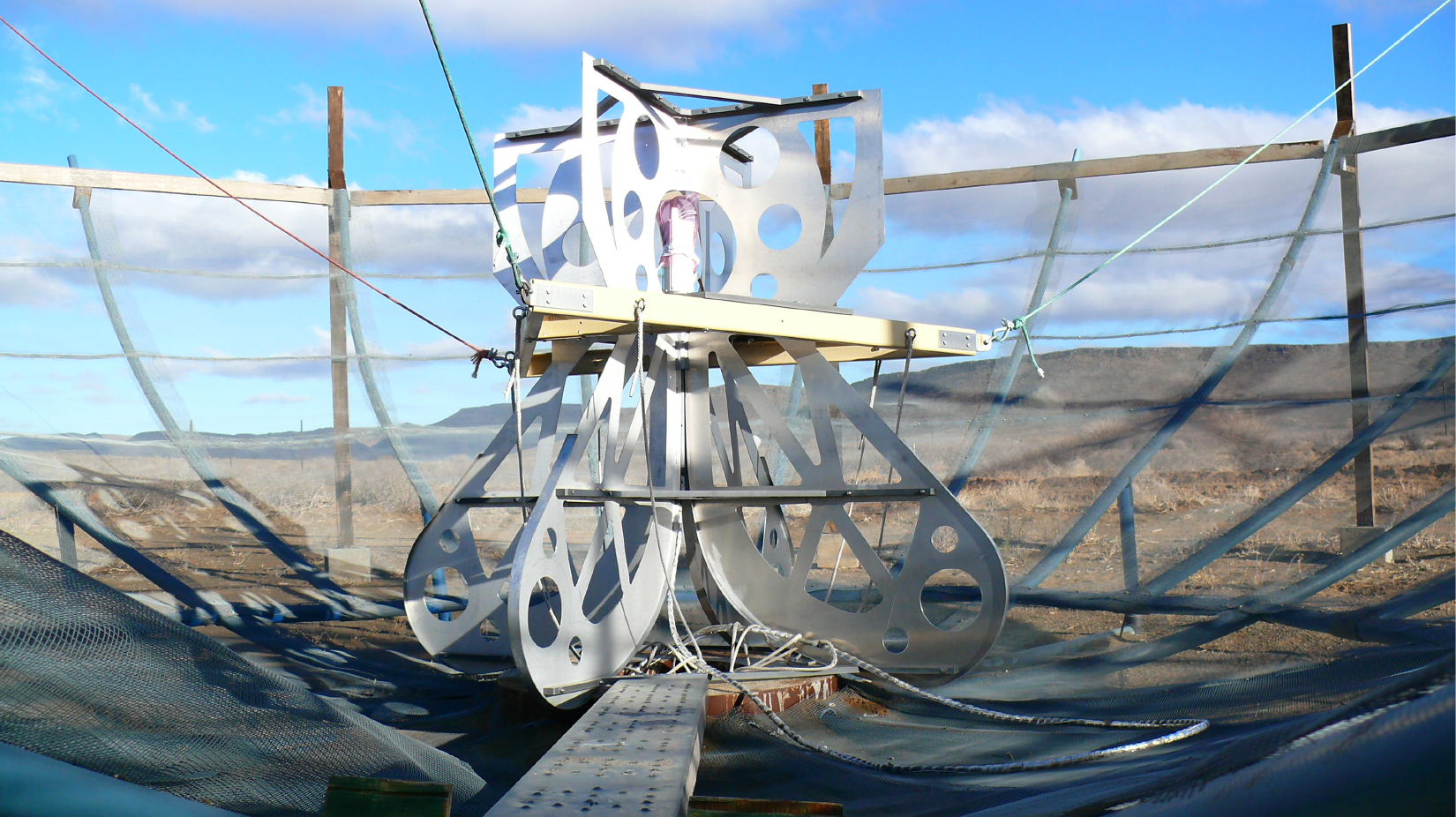}
    \caption{A lowered Vivaldi feed sitting in a 14-meter dish. The Front End Module (FEM) which acts as balun, LNA, and RFoF converter is the white rectangle mounted in the circular void within the top half of the feed. It is covered in a weather resistant bag to reduce environmental degradation.}
    \label{fig:feminfeed}
\end{figure}

The design goal of the signal chain is to minimize spectral structure as well as minimize response to out of band signals. Phase II replaced the analog signal chains with upgraded Front End Modules (FEMs) and Post Amplifier Modules (PAMs), and changed from 75 Ohm cables to optical fiber. The system is further described in \citealp{herarf}. The FEM is mounted on the Vivaldi feeds, and covered in a weather protective cover. This mounting can be seen in figure \ref{fig:feminfeed}, in a dish that has had its feed lowered for maintenance. The FEM takes in balanced, dual polarization signals from the feed, amplifies them and converts them to single ended 50 Ohm impedance. 

In addition to filtering and amplification components, the FEM includes a integrating Dicke switching radiometer \citep{dicke}. Between the first LNA and the second stage amplifier is a switch which can select a 50 Ohm load instead of the antenna. This load circuit also has a calibrated noise source that can be enabled. By measuring sky, load, and noise parameters in field,  the downstream portion of the signal chain can be absolutely calibrated. Integration of this calibration scheme is a work in progress. Currently, the load and noise settings are used often for commissioning and array quality assurance. 

The FEM also contains the initial 180 degree phase switch of a crosstalk mitigation system, based on the Walsh switching technique \citep{phaseswitch}. This system is currently being commissioned and has not been in use for previous seasons of HERA observing. In order to reduce spurious signals introduced along the analog signal chain, orthogonal Walsh functions are generated downstream on the signal processing boards and fed to the phase switch in the FEM over differential lines to enable synchronous switching of the signal chain phase. The phase of each polarization in the FEM can be controlled independently. 

Several digital devices in the FEM are included to provide useful diagnostic telemetry. These include sensors for voltage, current, magnetic heading, barometric altitude, and tilt. Control signals for all of these devices are sent via an I2C bus which is extended to long distances via a differential CAN bus transceiver over twisted pair lines in Cat7 ethernet cables. Wire pairs in this cable also transport the phase switching signal.

A notable change in the phase II design is the replacement of 35-meter coaxial cable for the connection between the FEM and PAM with a 500-meter RF over Fiber (RFoF) system. In order to minimize signal loss, the original coaxial cables were designed to be as short as possible while still allowing for multiple signals to be directed to a central node. However, the reflections in the coaxial cable introduced chromaticity at cosmologically interesting modes, and they were found to leak RF, which generated unwanted correlation between antennas. Therefore, the coaxial cables were passed over in favor of RFOF. A 500-meter length was chosen in order to place reflections well outside of delays of interest. With signals propagating at optical wavelengths there is no risk of unwanted re-radiation or mutual coupling within the cable runs. 

Cabling to the feed includes a pair of signal fibers, a coaxial power cable, and a Cat7 ethernet cable carrying digital monitoring and switching. Electromagnetic simulations suggested the bundle should be sheathed in a grounded cover but that contact of sheath with feed could result in additional chromatic structure.  The solution is an additional layer of weather wrapping and a carefully laced route through the antenna. Figure \ref{fig:leveling} shows wrapping for the cables extending to the bottom of the dish.

\begin{figure*}[ht]
\centering
\includegraphics[width=2\columnwidth]{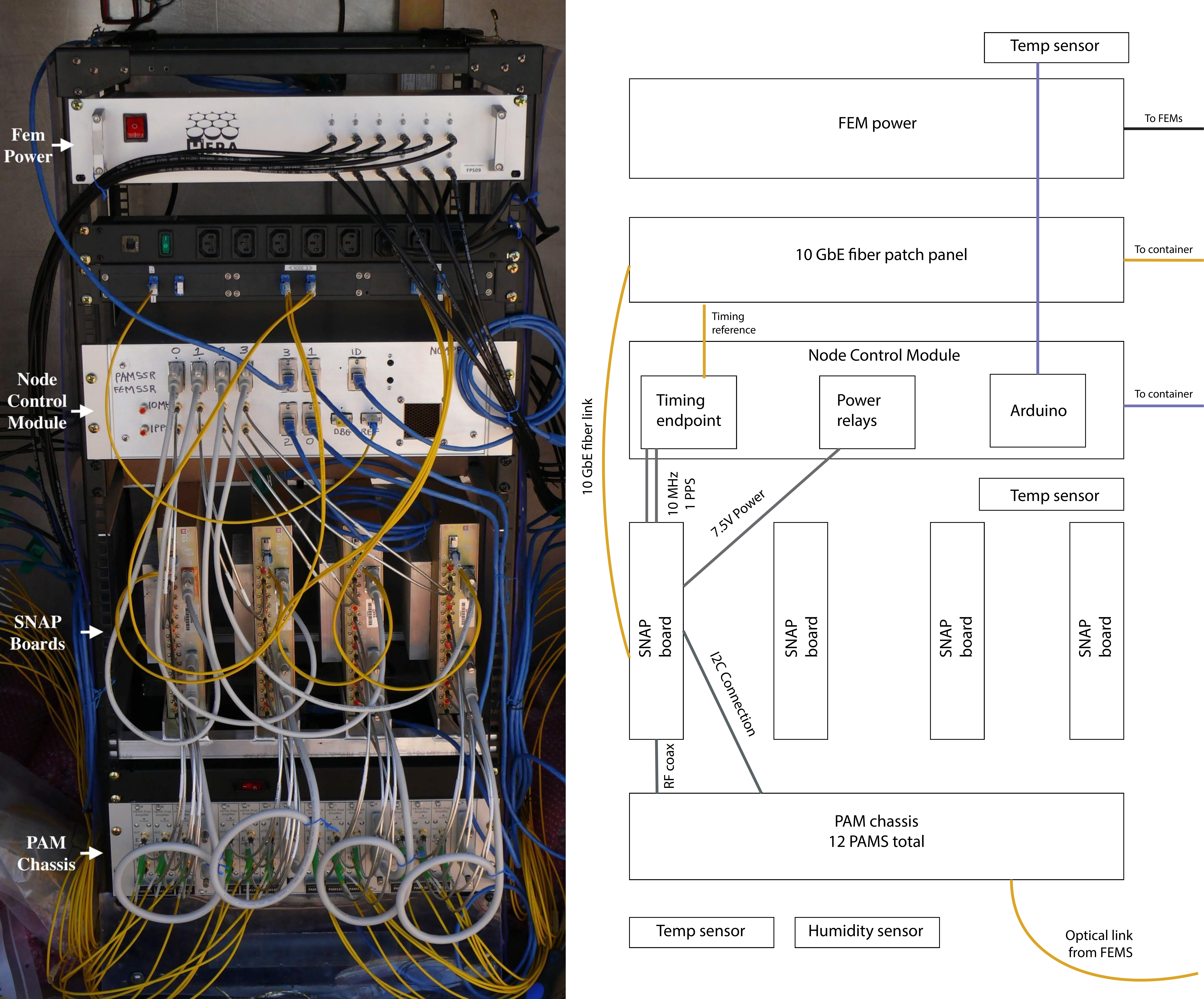}
\caption{ (left) The internals of a node which support up to 12 nearby antennas. (right) A diagram of the node. Cable colors correspond as closely as possible to those in the photograph. Bottom to top: RF over Fiber conversion in the PAMs, digitization and channelization in SNAPS, node control module supplies power switching and timing synchronization. At the top a patch panel links 10 and 1 GbE fibers back to the central switch bank. Cool air enters from ground loop beneath and exits at the top. Polarized signals travel in pairs. SNAPs also read out and control sensors and settings embedded in the signal chain.}
\label{fig:node}
\end{figure*}

At the node, a patch panel feeds the RFoF cables through to the PAM receivers. The node serves up to 12 antennas, and contains part of the analog signal chain, signal processing, and digital control systems. Figure \ref{fig:node} shows a photo of the inside of the node along with a schematic view. The node structure is an RFI tight enclosure which is cooled by air forced through a buried ground loop. The PAM receiver chassis can be seen at the bottom of the node. Additional node components will be discussed in their respective sections.

Each PAM receiver converts both polarizations of one antenna back to RF. It applies gain and filtering of those signals, acting as the anti-aliasing filter for the analog to digital converter (ADC). Each PAM also has a digitally controllable attenuator, which can be set independently for each polarization, and allows for the input signals to the ADC to be levelled by up to 14 dB. The PAM sits on the same I2C network as the FEM and provides a physical connection point for the digital cable to the FEM.

\begin{figure}[ht]
    \centering
    \includegraphics[width=\columnwidth]{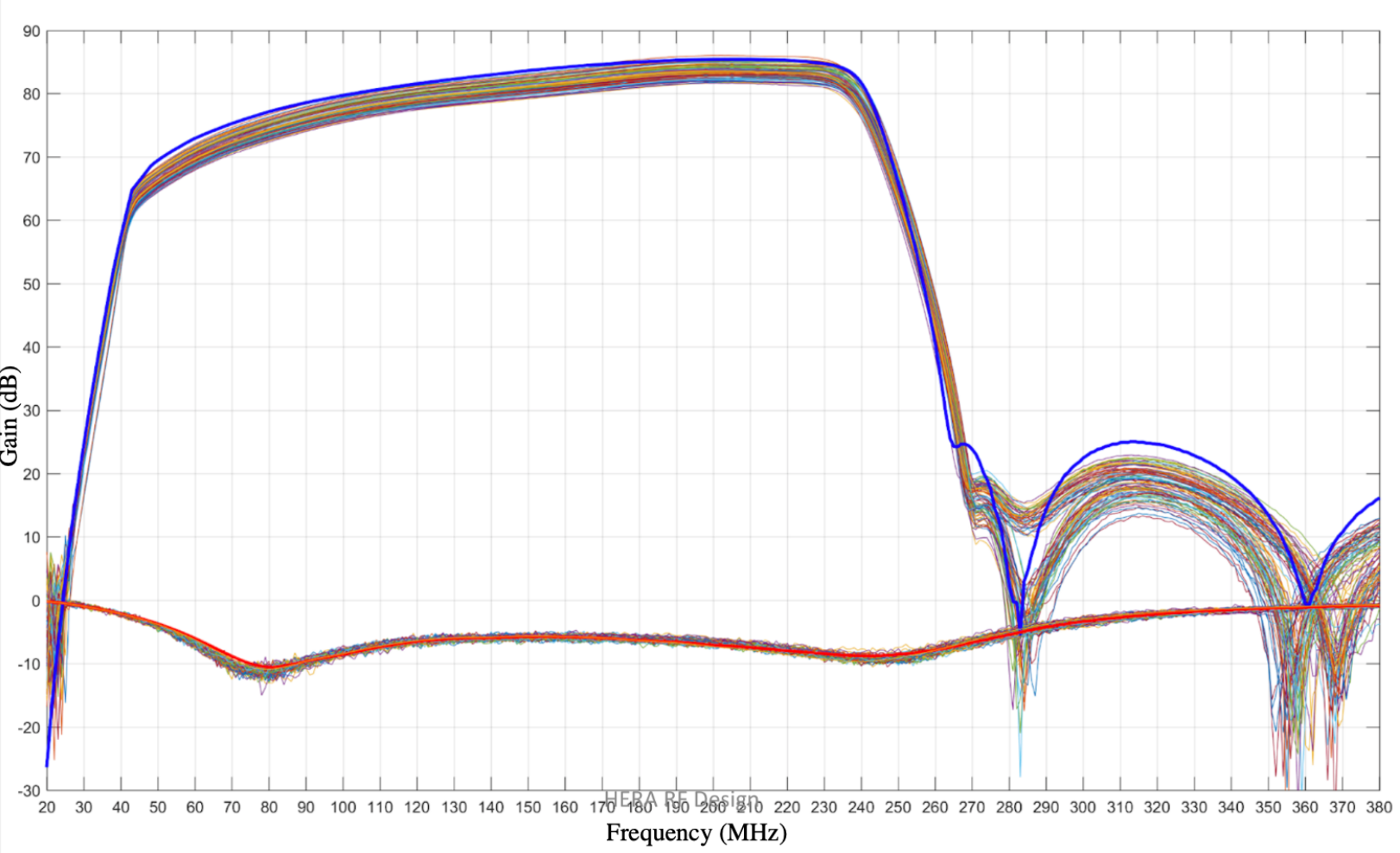}
    \caption{3-port VNA measurements of the FEM - PAM system, as compared with simulation values. The simulated S21 measurement is highlighted in blue, and the S11 measurement in red. 24 lab modules are measured for comparison. These modules were designed to introduce minimal extra chromaticity in the 50-250MHz passband. The total gain of the analog system in the middle of the band is approximately 80 dB.}
    \label{fig:VNA}
\end{figure}

Figure \ref{fig:VNA} compares the expected gain spectrum of the analog chain with several modules measured in the lab. The S21 measurements with a two port VNA show the throughput gain of the system, with the simulated data highlighted in blue. The full gain of the analog system measures around 70 dB at the low end of the band, and 85 dB at the high end. The noise figure is approximately 90k in the middle of the band. To approximate the sky temperature for a ``quiet'' patch of sky away from the Galactic plane, we can use a power law \citep{Furlanetto_2016}:
\begin{equation}
    T_{sky} \approx 180 \left( \frac{\nu}{180 \text{MHz}}\right) ^{-2.6}\text{K}
\end{equation}

We recover a sky temperature of $\approx$ 290k at 150 MHz.

The S21 response is suitably spectrally smooth within the 50-250 MHz bandwidth. The S11 measurements show the power reflected from the analog system, with the simulation data highlighted in red. The reflected power is also minimally chromatic within the HERA band. 

\subsection{The Correlator} \label{sec:correlator}
The correlator system digitizes, channelizes, and cross correlates between all antennas. The system begins at digitization and ends at file writing. To minimize the number of components the correlator system is also responsible for ancillary services like RF signal chain telemetry, timing, and networking.  The correlator uses the FX architecture, where inputs are channelized first and cross multiplied second, and has a scalable design based on the framework described in \citealp{Parsons_2008}. 

The system is distributed across multiple embedded and traditional compute systems.  Settings and status values are synchronized via a REDIS\footnote{https://redis.io/docs/} in-memory data store. REDIS captures the current state of the correlator and related systems, but does not contain any history. REDIS is polled regularly and the current state is timestamped and stored in a database with historical capabilities, described further in section \ref{sec:mc}.

\subsubsection{F-engine}
\begin{figure}[ht]
    \centering
    \includegraphics[width=\columnwidth]{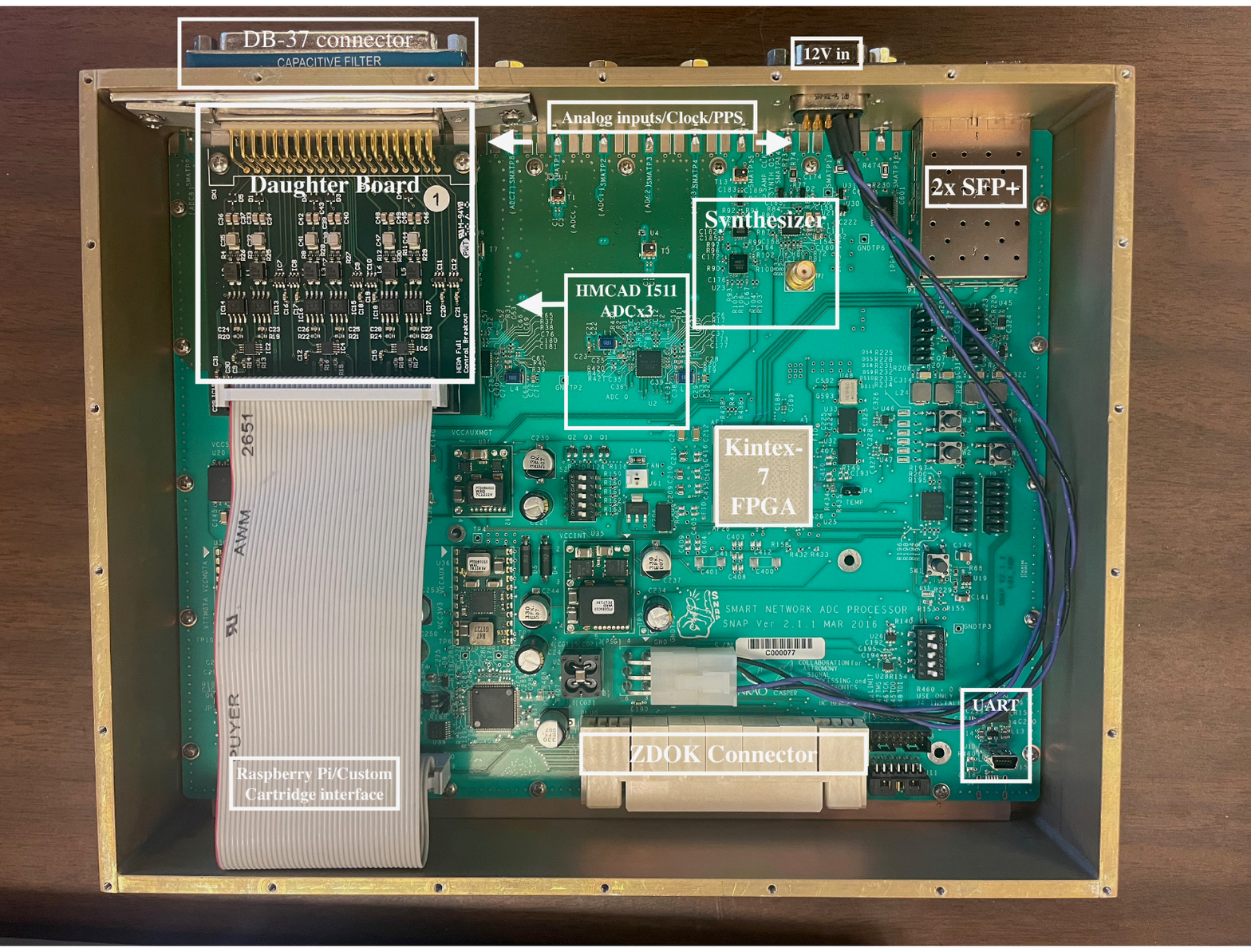}
    \caption{Top-down view of a Smart Network ADC Processor (SNAP) board in its chassis. The daughter board is connected through the Raspberry Pi/custom card interface. 12 analog inputs are available, though due to the necessary ADC processing speed, only 6 are used for HERA. SMA inputs are also available for the reference clock and PPS timing signals, which can be up-converted by the onboard synthesizer. The SNAP has 3 onboard HMCAD1511 ADCs, as well as a Kintex 7 FPGA. Three interfaces on the back of the board allow for external connections, over a ZDOK, UART connector, or Raspberry Pi interface. HERA uses the Raspberry Pi interface to mount the digital control daughter board, but does not use the other interfaces. The SNAP can be controlled via a Raspberry Pi processor, but the HERA firmware substitutes a microblaze softcore processor.}
    \label{fig:SNAP}
\end{figure}

The phase II array upgrades the signal processing boards from the Roach II boards and 16 input ADCs used by PAPER to a Smart Network ADC Processor (SNAP) board designed by the Collaboration for Astronomy Signal Processing and Electronics Research (CASPER) \citep{casper} in collaboration with the NRAO. A top-down view of the SNAP is shown in figure \ref{fig:SNAP}. Input signals are digitized by HMCAD1511 ADCs and read out by a Kintex 7 FPGA. The board is controlled by a microblaze softcore processor on the FPGA that is accessible via a Python programming interface. The SNAP acts as both the digitizer and the ``F-engine'', taking in analog signals over coaxial cable and outputting Fourier-transformed spectra over 10 GbE fiber. Each SNAP reads out 3 dual polarization antennas (six signal chains) and samples them at 500 Msps for a 250 MHz bandwidth. In order to achieve this speed, the ADCs are run in an interleaving mode. Each ADC device has four digitizer channels with two interleaved channels digitizing each signal. The SNAP is clocked from a 10 MHz reference and a 1 PPS timing signal generated by a timing endpoint within the node control module. The endpoint used relies on the White Eabbit technology \citep{whiterabbit}, developed partially by CERN. The 10 MHz reference is converted by the on-board synthesizer to a 500 MHz clock for the ADCs and a 250 MHz clock for the FPGA.

\begin{figure*}
    \centering
    \includegraphics[width=1.8\columnwidth]{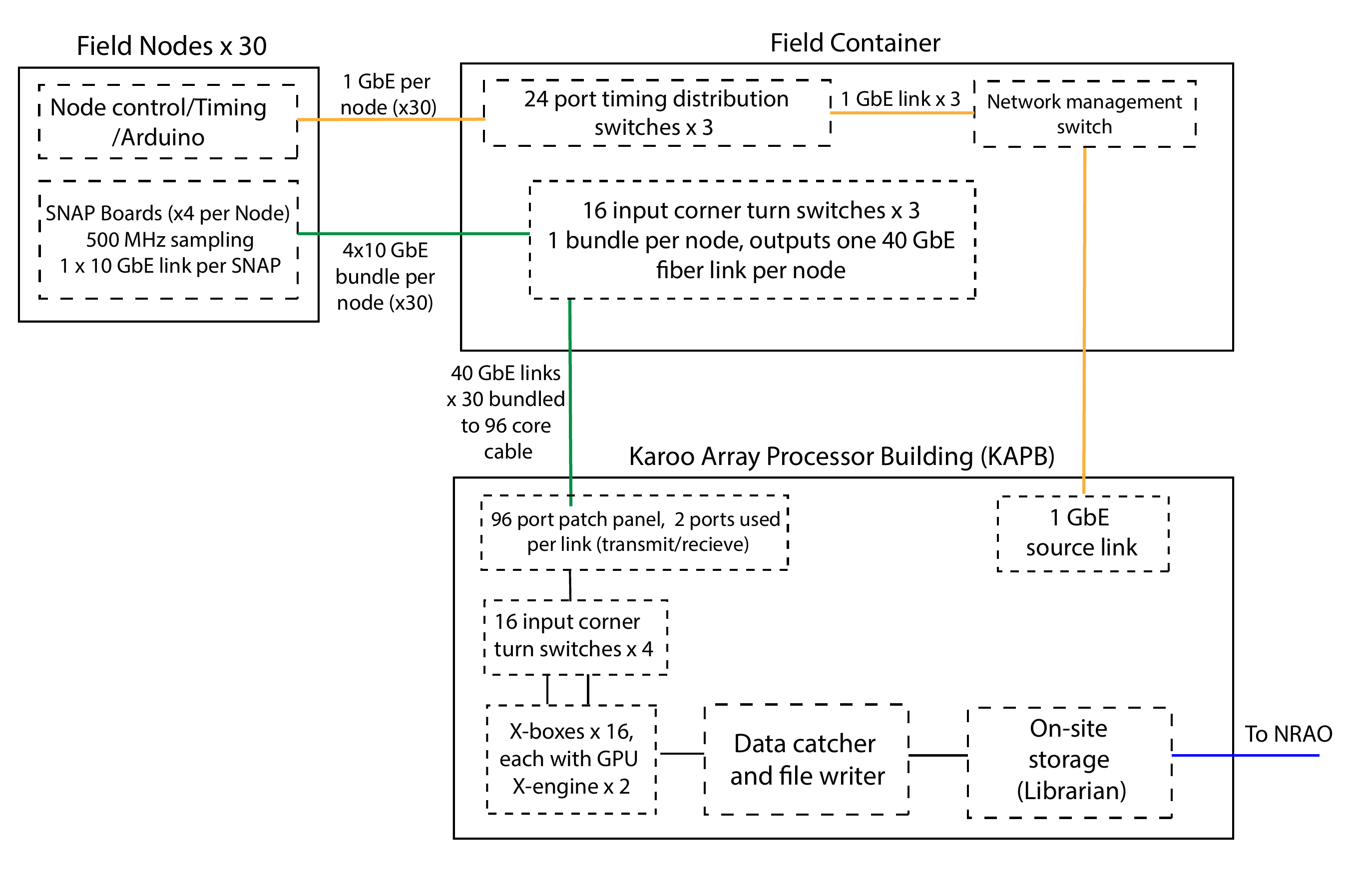}
    \caption{An overview of the networking scheme. The field nodes send out packets over 10 GbE fiber. These connect to  ``corner turn'' switches in the field container, which send antennas to the GPU X-engines. These fibers are then bundled and sent over a 10 km link to the KAPB, where a second set of switches completes the corner turn transpose. This allows the NIC and GPU in the X-box to perform parallelized cross correlation. The outputs of the correlator are averaged and sent to a dedicated catcher machine which writes raw visibility files. These are then converted to UVH5 format, uploaded to the on-site Librarian (described in section \ref{sec:librarian}), and then synced to NRAO. Yellow connections indicate a 1 GbE link, green indicates 40 GbE transport, blue is the Globus 200 Mbps link to NRAO.}
    \label{fig:networking}
\end{figure*}

While these boards could be mounted in the node directly, to limit crosstalk and self interference each board is encased in a custom RFI tight chassis. The chassis also acts as a passive thermal control system conducting thermal energy from the FPGA into the aluminum case.  Four of these modules are mounted in each node as shown in Figure \ref{fig:node}. The SNAP outputs spectrum packets over a 10 GbE SFP+ connector, visible in the figure as yellow fiber connections.

The FPGA design contains a polyphase filter bank, Fourier transform, equalization, as well as a number of other useful functions, including an onboard correlator for rapid testing and prototyping of the system. Additionally, the Walsh patterns for phase switching are generated on the SNAP and converted to differential signals by a custom daughter card for off board communication to PAMs and FEMs. 

The F engine channelizes to $2^{13}$ (8192) channels for a resolution of 30kHz. However due to bandwidth limitations in the 10 GbE connection, only 178.5MHz of bandwidth (6252 channels) is sent. 

\subsubsection{Networking and X-engine}
Spectrum packets from a node's four SNAPs are sent over 10 GbE ethernet to one of 3 10/40 GbE Ethernet switches in a shielded container adjacent to the array. These switches form part of the the ``corner turn'' data transpose which rebroadcasts antennas to multiple GPU ``X-engines'' for cross multiplication. From here these switches are crosslinked over a 10 km fiber bundle to the Karoo Array Processor Building (KAPB) and a matching set of  switches, which complete the corner turn transpose and send the reordered data to the GPU X-engines. The KAPB also hosts all other HERA on-site compute including monitoring and on-site data archiving.

Cross multiplication is performed by 32 total GPU-based X-engines, housed two to a server, with each GPU allocated a 10 GbE Network Interface Card (NIC). Each X-engine receives 1/16th of the spectra from every antenna with half the engines reading in the even time samples and the other half odd. The engine first averages by a factor of four to 122 kHz channels, cross multiplies all pairs of antennas, and every 8 seconds sends an averaged visibility to a dedicated catcher machine which saves data to 
disk. An overview of the networking setup is shown in figure \ref{fig:networking}.

\subsubsection{Data Catching and Compression}
 Interferometers with a large number of antennas are constrained by data volume and network bandwidth concerns. The connection between the Karoo site and NRAO has an average 200 Mbps transfer bandwidth, and the on-site storage availability is approximately 1.1 Petabytes. Reducing data volume is a high priority concern, especially as HERA builds out to the full 350 antennas. Averaging data in frequency or time can reduce the data size, but can cause smearing effects that degrade instrument performance particularly on the longest baselines.

Data are written in a two step process to minimize the impact of time spent applying compression on data recording. The correlator writes a raw custom format with minimal metadata information. Once written these are converted to an HDF5 format called UVH5 readable by the \texttt{pyuvdata} \footnote{\url{https://github.com/RadioAstronomySoftwareGroup/pyuvdata}} \citep{Hazelton2017} library with compression using the Bitshuffle algorithm \citep{bitshuffle}, originally developed for the CHIME telescope, to reduce the volume. This method is independent of time or frequency averaging, and can be applied alongside such schemes. In practice, for HERA this offers a 2:1 compression ratio. 

One avenue under exploration to further reduce the data output of the telescope is baseline-dependent averaging (BDA). BDA has been investigated in the context of the SKA \citep{bda} and is expected to greatly reduce visibility data sizes with minimal loss to instrument performance. As the shortest baselines in the array do not require short integration times and fine channel resolution to limit smearing and decorrelation effects, these baselines can be averaged for longer and in broader frequency chunks, reducing the data volume. In the context of EoR science with HERA, we will only target time domain BDA, as the EoR measurement is very sensitive to spectral features and we do not want to risk introducing frequency dependent structure. As the majority of HERA's baselines are relatively short, this solution has the potential to greatly reduce data volume. 

The averaging time set for each a baseline can be set to match the amount of allowed decorrelation experienced. The decorrelation is highest for angles corresponding to high fringe rates; generally large angles with sources far from zenith \citep{bridleandschwab}.

For example we might allow, for a source 10 degrees off zenith, a maximum decorrelation of 10\%, and calculate the maximum allowed integration times for each baseline. Realistically, there are also a few extra limitations imposed. Sources moving across the stationary beam lead to flux changes with time. Averaging these sources over long periods of time will introduce uncertainty in the source amplitude. Practically this limitation amounts to a few minutes for our reference sky position. Additionally, the size of buffers in the GPU X-engine impose a roughly 30 second limit on integration time. Combining this all together gives a potential compression factor of 12.

However BDA does require adapting a long standing assumption of uniform integration time in many downstream data analysis steps. As a practical first step in this direction, 
 data in the 2023 season will be split into two files, one averaged suitable for core baselines and a second faster cadence including outrigger baselines to prevent decoherence. This will allow for testing of the BDA infrastructure in expectation of a full future deployment.

\subsection{Monitoring and Data Storage}
The collection of systems which make up HERA adds up to hundreds of computers, switches, and sensors. Detailed monitoring and data management is essential to obtaining high quality data across 8 months of continuous observing. Applying the same logic, data quality must be assessed in near real time with a first round of analysis. Ancillary systems for tracking, cabling, configuration, and managing the data archive have all been designed with the goal of maximizing the uptime and quality of every antenna. 

\subsubsection{Configuration Management}

At 350 dishes there are many connections to keep track of. Careful record keeping is essential to reconstruct the full state of the system at any point in HERA's lifetime. This also minimizes telescope down time due to miscabling or mislabelling. Similar systems are in place on CHIME \citep{chimecm} and other large N interferometers.

Each antenna connects through nine RF interconnect points and more than 8 digital interconnects. The Configuration Management (CM) system tracks these connections and provides this information to downstream systems to map input numbers to antenna numbers and provide data in a ready-to-use form. 

Real time analysis products use part connection information to display data sorted according to location by node, SNAP, etc. This is extremely useful for diagnosing systematics due to part malfunction.

In addition to tracking part connections, CM stores the location, serial number, and history of antenna parts for the purposes of inventory and identifying long term failure modes.  

\subsubsection{Digital Control} \label{section: digcontrol}
\begin{figure*}[ht]
    \centering
    \includegraphics[width=2\columnwidth]{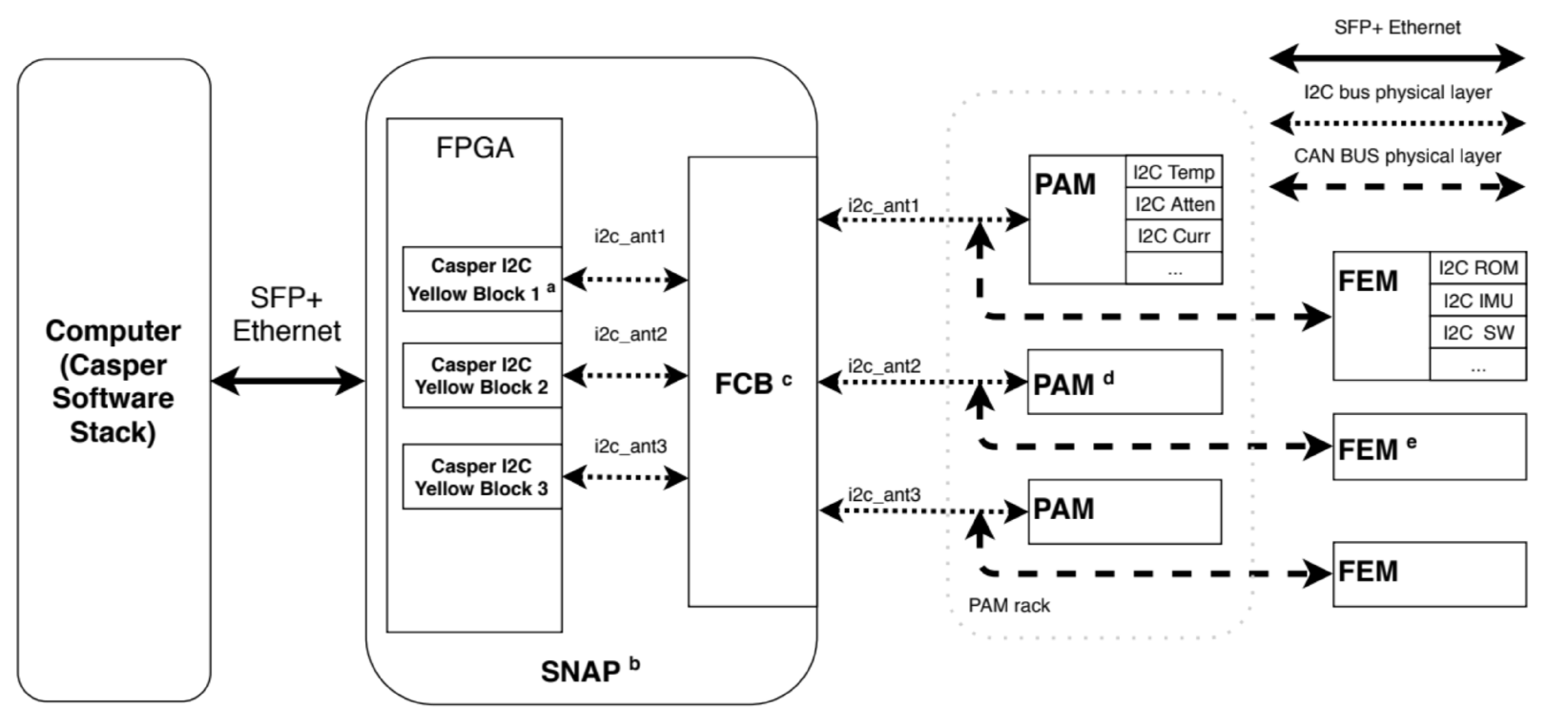}
    \caption{The digital control scheme. Software running on a control computer uses the CASPER Python API to communicate with I2C blocks on the FPGA firmware. These act as the primary controller on a network which extends, via the daughter differential transceiver board (FCB) to the PAMs and FEMS. The PAM and FEM communicate over a CAN BUS physical layer. This allows for commands to be passed from the computer through to the antenna front end, for controlling switches in the FEM, as well as allowing telemetry data to be communicated back to the host computer.}
    \label{fig:digcontrol}
\end{figure*}

The correlator system includes digitization and cross correlation, but also ancillary elements related to the node health.
Figure \ref{fig:digcontrol} shows the control scheme. The SNAP board provides digital logic link to the digital components in the RF signal chain. These, plus the F-engine itself are controlled and monitored via  CASPER software in Python on a dedicated computer on the SNAP network. 


 Environmental, power and White Rabbit timing services in a node are located in the control module (NCM), which can be seen in figure \ref{fig:node}. The NCM contains an Arduino microprocessor which operates sensors and power relays and a central server that sends and receive microprocessor messages and reports to a database. This system controls power to SNAPs, PAMs, and FEMs and monitors temperature and humidity within the node.  Temperature sensors are placed at high, mid, and low locations within the node to monitor heat flow through the node stack.

\subsubsection{System Monitoring} \label{sec:mc}

HERA is a large distributed set of systems which each provide logs, measurements, and system checks to a central Monitor and Control system. Many of these systems also accept commands or configuration information, or are configured by hand in a way that must be tracked.  The M\&C system is designed to be able to fully reconstruct the historical state of the system, to allow the telescope team to trace failures, and to allow data analysts to trace systematics in historical data. It acheives this goal using a central database which is accessible via an application layer Python package called \texttt{hera\_mc}\footnote{\url{https://github.com/HERA-Team/hera_mc}}. Since the system touches every other system, it must also be reliable and fault tolerant so it does not limit the uptime of the telescope.

Data stored in M\&C comes from a wide variety of subsystems that include the logging from the correlator, node control module, analog components, the site weather station,  Librarian, RTP, and compute state reporter (memory, uptime, ip address, raid errors). 

These data are collected by either polling subsystems through their provided interface or via push from the subsystem itself using the M\&C software interface.  Information is stored in an Postgresql database, which can be read in many different ways. The \texttt{hera\_mc} package offers a Python API built on SQLAlchemy \citep{sqlalchemy}, and version controls the database schema using the Alembic database migration tool. All items are tagged with GPS time (seconds since Jan 1 1980). This time standard was chosen over unix epoch for robustness across leap seconds and over Julian date which requires large floating point precision to be accurate to the second.    

The M\&C database is mirrored as ``hot standby'' to several off site locations including the NRAO where it can be used for analysis and to a web server for display. The database is also regularly backed up on-site. The web backup is used to update live dashboards described in section \ref{dashboards}.

\subsubsection{On-site Real Time Processing} \label{RTP}
Thousands of observing hours with many antennas are necessary to reach HERA's desired sensitivity level. Obtaining this much data requires continuous observing through a seasonal 8 month long campaign. Storms,  interference, computer failure, and cable breaks do occur. The more subtle system failures are not necessarily detected in telemetry monitoring but are obvious during data reduction as an increase in noise, a calibration error, or other flag triggers.  For this reason we have increasingly moved the best understood parts of the data reduction pipeline into an on-site analysis step which runs on a nightly basis.

The Real Time Processor (RTP) system is a collection of steps including calibration and flagging executed by a customized automated pipeline infrastructure. The infrastructure system used to run the pipeline is well described in \citealp{La_Plante_2021}. The compute resources are a relatively small cluster of 8 machines with common access to recently captured data in a Network File System (NFS) and indirect access to the a large archive (the ``Librarian'' described in section \ref{sec:librarian})  via rsync. Like the other HERA systems running on-site, it runs unattended with limited remote operator oversight. This requirement places a premium on automation and redundancy. The resulting analysis is posted nightly to a web server for inspection by observers.

\begin{figure}[ht]
    \centering
    \includegraphics[width=\columnwidth]{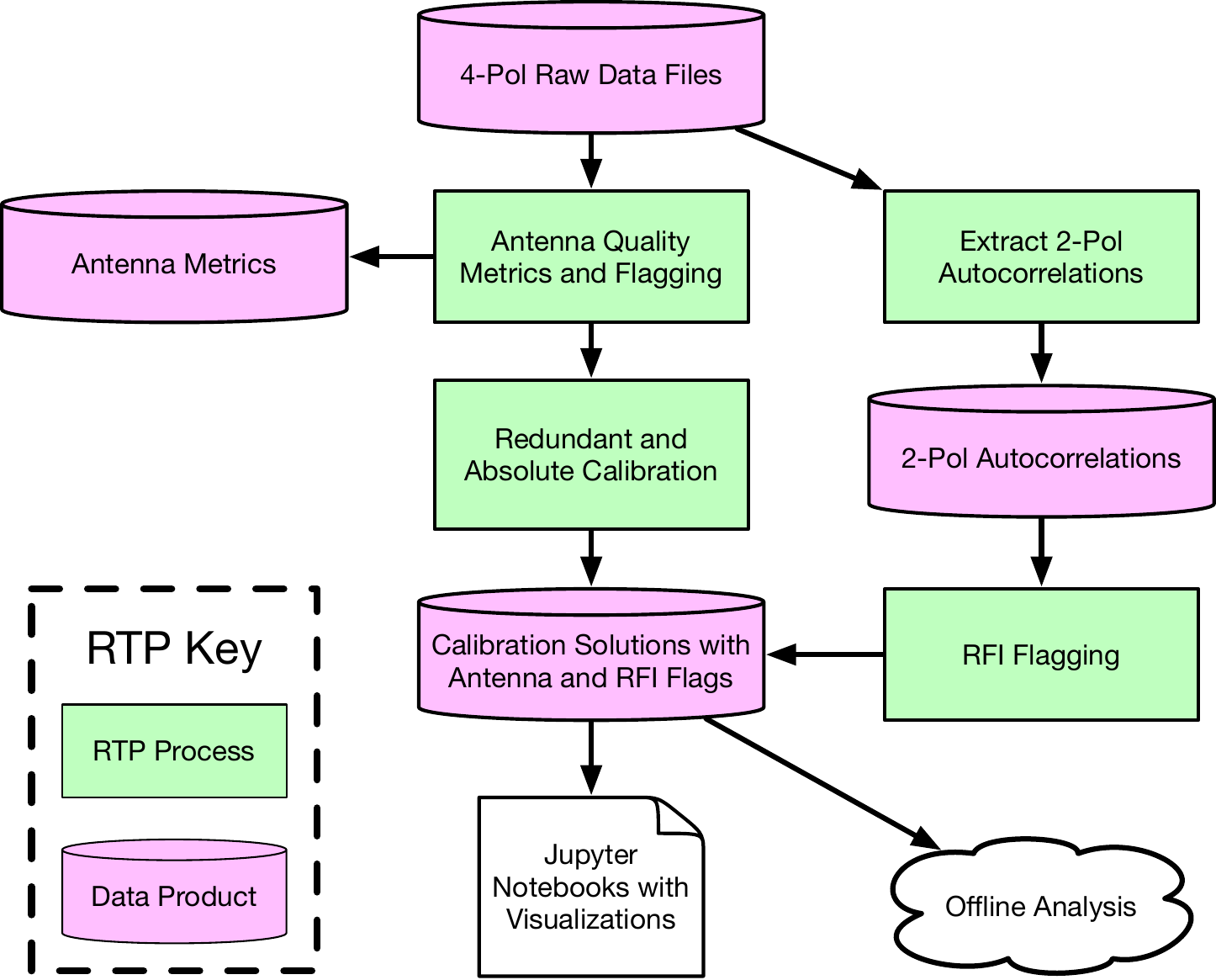}
    \caption{A simplified view of the flow of analysis steps applied in near-real time by RTP. The key tasks are antenna and RFI flagging and calibration, all of which are useful for rapid assessment of overall data quality and array health. Further processing is done offline at NRAO.}
    \label{fig:rtpflow}
\end{figure}

The on-site pipeline steps include: flagging, calibration, filtering, and smoothing. It separates out autocorrelations into their own files, produces calibration solutions using both redundant calibration and a calibration model, as well as derives various quality metrics. The flow of these steps is shown in figure~\ref{fig:rtpflow}. Summary plots for inspection are output as Jupyter Notebooks. These are discussed in section \ref{sec:notebooks}. 

\begin{figure*}
    \centering
    \includegraphics[width=1.8\columnwidth]{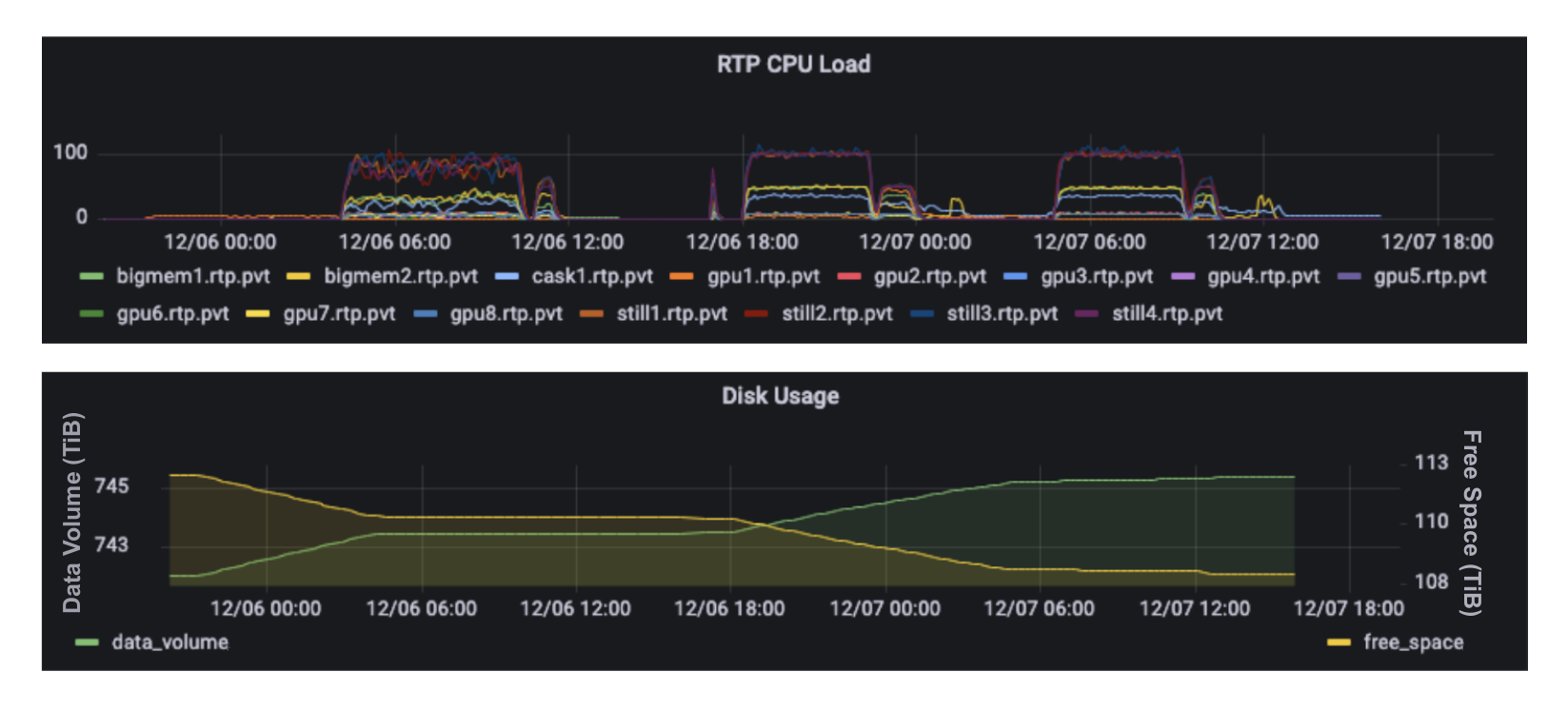}
    \caption{(top) 
    Pipeline computer loads captured by the monitor and control system illustrate a sustainable processing schedule over a few observing days. (bottom) Data volume vs. free space for a few observing days. First, for this observing season, data are captured at a rate of approximately 0.3 TiB/hour, seen here as a decline in free space and increase in data volume in the bottom panel. Once observing finishes in the morning, processing begins, seen here as sustained CPU load in the top panel. Then at 0600 the second night observing completes and the process repeats. At this time, 18 hours time lapsed between compute cycles so a backlog night was added at 1800 hours.}
    \label{fig:diskusage}
\end{figure*}

RTP launches automatically at the end of data recording every night starting with the raw data files written by the correlator and to stay current must complete before the resumption of observing the next day. Figure \ref{fig:diskusage} shows a typical nightly cycle of resource usage in the on-site cluster. Makeflow is used to construct a logical execution order which takes into account dependencies between tasks. A custom wrapper around Makeflow, called \texttt{hera\_opm}\footnote{\url{github.com\\HERA-Team\\hera_opm}} (HERA online processing module), applies rules devised specific to HERA processing such as ``time chunking'' of data which batches together analysis steps like RFI excision which require multiple time samples. Makeflow executes new tasks once previous tasks have produced any pre-requisite inputs. For example, notebooks which plot an entire nights-worth of calibration solutions must wait for all files to be completed. Makeflow executes tasks using Slurm \citep{Slurm} which assigns jobs to compute resources according to available capacity and the requirements of the job.

The current storage available on-site is approximately 1.1 Petabytes distributed across seven servers. Each server hosts a set of disks in the ``RAID 60'' configuration with three ``hot spare'' disks that can be used by any individual RAID 6 partitions. This configuration  incurs approximately a 25\% penalty of data storage capacity compared to the raw disk size, but allows for at least 5 individual drive failures before data is lost.

\begin{figure}[ht]
    \centering
    \includegraphics[width=\columnwidth]{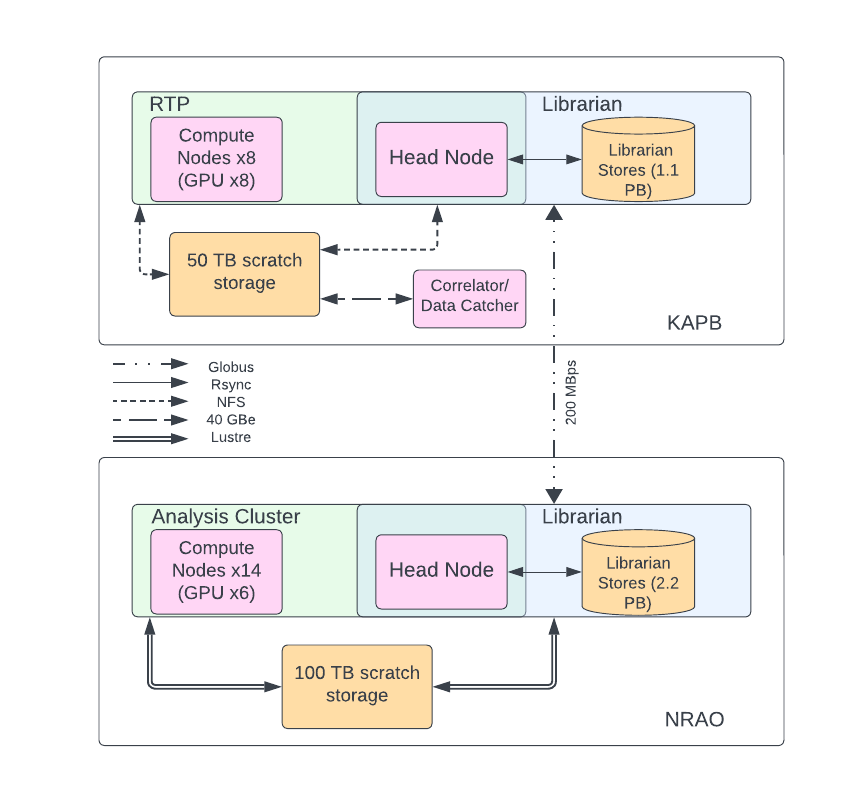}
    \caption{The archiving and processing architecture on and off site are functionally mirrors. The real time processing (RTP) system executes complex pipelines while the Librarian system manages and provides an interface to the data archive. The on-site pipeline calibrates and flags a nights worth of data starting when observing finishes for the night and ending before observing starts again the next night. The Librarian manages the $\sim$1.1 PB of on-site storage and transfers data off to the $\sim$2.2 PB storage at the NRAO using a database to track associated analysis products and to safely delete data once it has been safely moved off site.}
    \label{fig:rtp}
\end{figure}

Raw data and pipeline products are streamed for offsite backup and further analysis to the NRAO data center in Socorro, NM.
Figure \ref{fig:rtp} shows an overview of the RTP on-site and off-site systems, and how they interact. In the on-site cluster, the raw data storage can be accessed by the compute nodes and the head node for processing, and is also uploaded to the on-site Librarian stores. The connection between the Karoo site and NRAO has an average 200 Mbps transfer bandwidth and a typical latency of 100 to 500ms. Data is transferred to the off-site cluster at NRAO via rsync or Globus \citep{globus} depending on service availability. Globus uses a third party coordinator to manage parallel UDP connections to significantly increase transfer speeds over high latency connections.  

In addition to reports presented to observers, the near real time returns from nightly calibration and flagging are also stored in M\&C (see section \ref{sec:mc}) for analysis of long term trends which can be tracked in a dashboard system discussed further in section \ref{dashboards}.

\subsubsection{Data Archives} \label{sec:librarian}
 A typical nights observation generates 1800 raw files and RTP processing generates of order ten more per file. These files are stored across several disk servers, where they can be quickly found and retrieved. This set of requirements was found to need a custom solution. The Librarian \citep{La_Plante_2021} was developed to meet these unique needs. Each Librarian instance consists of several independent RAID server nodes and a head running the API server and database. One instance is located at the telescope and another at the primary archive where data are stored for long term. Additional sites have been operated for local use as necessary.
 
 Data storage requirements have increased as antennas have been added with each season. The capture rate in the 2023 season is forecast to be  0.454 TiB per hour, assuming 256 operating antennas with the current settings of 8 seconds averaging time and 2048 channels. Over an entire 2000 hour observing season, this will amount to almost 850 TB of data. The data must be synced across multiple sites for end-product users, and must be easily filtered using metadata. Additionally, since there is not enough storage on-site for the lifetime of the project, data must be regularly deleted once it is transferred off site. Deletion must be done only after verification of successful transfer. File locations and status are logged in a  PSQL database while a service program provides functions for uploading, transferring, and safely deleting copies of files. The server is accessed through an application programming interface (API) which can be used by automated analysis pipelines to get and put data as well as a web-based front end for human management. 

The primary functions of the Librarian are data ingest, location tracking, and transfer. On a nightly basis, the raw correlator data is output to a temporary buffer disk. It is then converted to the UVH5 file type and uploaded to the Librarian server, which creates a staging area for the upload and generates a one-use rsync command to be run by the uploader process. Additionally, also on a nightly basis, data are transferred via Globus from the on-site Librarian to the NRAO. Figure \ref{fig:rtp} shows the architecture of this transfer, as well as how it interconnects to the RTP. On the remote or on-site servers, users can search for this data to stage to their own working area. Using an SQL query through a web interface, the user can specify on a number of search parameters (such as unique file name, observation ID, Julian Date range, etc...) to find all files associated with the search. They can then copy those files locally for analysis. 

The telescope Librarian instance is co-located with the X-engines and real time processing cluster where it can ingest data at the highest possible rate. From there data are moved to the primary long term storage instance at National Radio Astronomy Observatory (NRAO) in Socorro, NM, which also supports post-processing analysis.  The processing cluster at NRAO has fourteen compute nodes with 8-core processors and 256GB of RAM.  The compute nodes have no local disk storage, but they are connected to a distributed parallel filesystem (Lustre) via a 40Gbit Infiniband network. Data can be staged from the archive to Lustre via the web portal or API.

\begin{figure*}[ht]
    \centering
    \includegraphics[width=1.8\columnwidth]{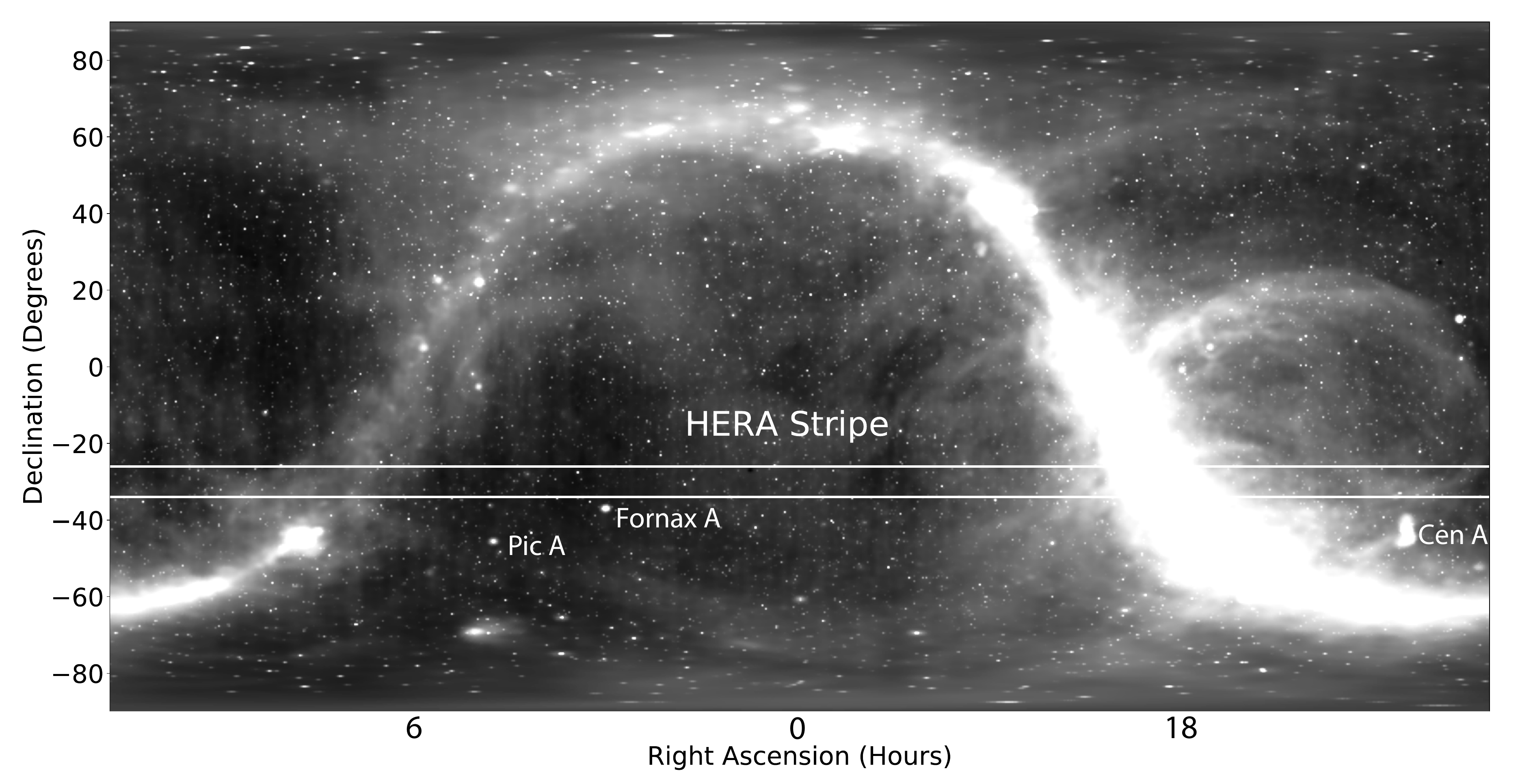}
    \caption{HERA's observing stripe, 8 degrees wide centered at approximately -30 degrees in declination. A few nearby bright radio sources are labelled. The background image is the radio sky including point sources and diffuse emission, combining data from the GSM \citep{de_Oliveira_Costa_2008}, NRAO VLA sky survey \citep{nvss}, and the Sydney University Molonglo Sky Survey \citep{sums}. Diffuse and point sources are not on the same scale, and are relatively scaled for visual clarity.}
    \label{fig:stripe}
\end{figure*}

\subsection{Observing}
The cosmological observing program must maximize time spent on the coldest foreground regions of the sky while allowing sufficient down-time for maintenance and improvements. The array location and drift scan design makes an observable sky region of a 8$\deg$-wide scan at -30d declination. This stripe is visualized in figure \ref{fig:stripe}. This stripe crosses both the bright Galactic center, and south alactic pole where diffuse foregrounds are dimmest. Due to bright and highly variable solar radiation in the radio band, as well as limited cooling capacity, observations are limited to the night time. The south Galactic pole is up at night in the Austral spring and summer. The combination of these factors result in an observing season beginning in August and lasting until April. During this time, sensitivity is maximized by recording with the largest number of antennas on the most hours possible. A continuous program of observing is required.

\subsubsection{Observers}

During science observing all points of the system from server health to data quality are monitored by remote observers.  Observing nightly across a season stretching over eight months is a continuous effort by volunteers from across the collaboration. 

The HERA collaboration members sign up for weeklong observing duty, with up to three observers per week. HERA observers have 3 primary goals. The first is to ensure automated observing started correctly and monitor the data capture process for disruption. The second is to make a rough judgement of data quality. This determination reflects major issues such as a known failure of a correlator component or a large number of unusual autocorrelation spectra. Lastly, observers report on whether the pipeline has run successfully and whether data is transferring to the off site archive. 

A large number of observers have volunteered to help maintain a 200+ day campaign of daily observing. Many are students or have expertise in other areas. Observing instructions and displays have been designed to require minimal expertise. All monitoring systems are accessed through a web browser. No software installation or training in data access is required.

\subsubsection{Dashboards}\label{dashboards}
\begin{figure*}[ht]
    \centering
    \includegraphics[width=1.8\columnwidth]{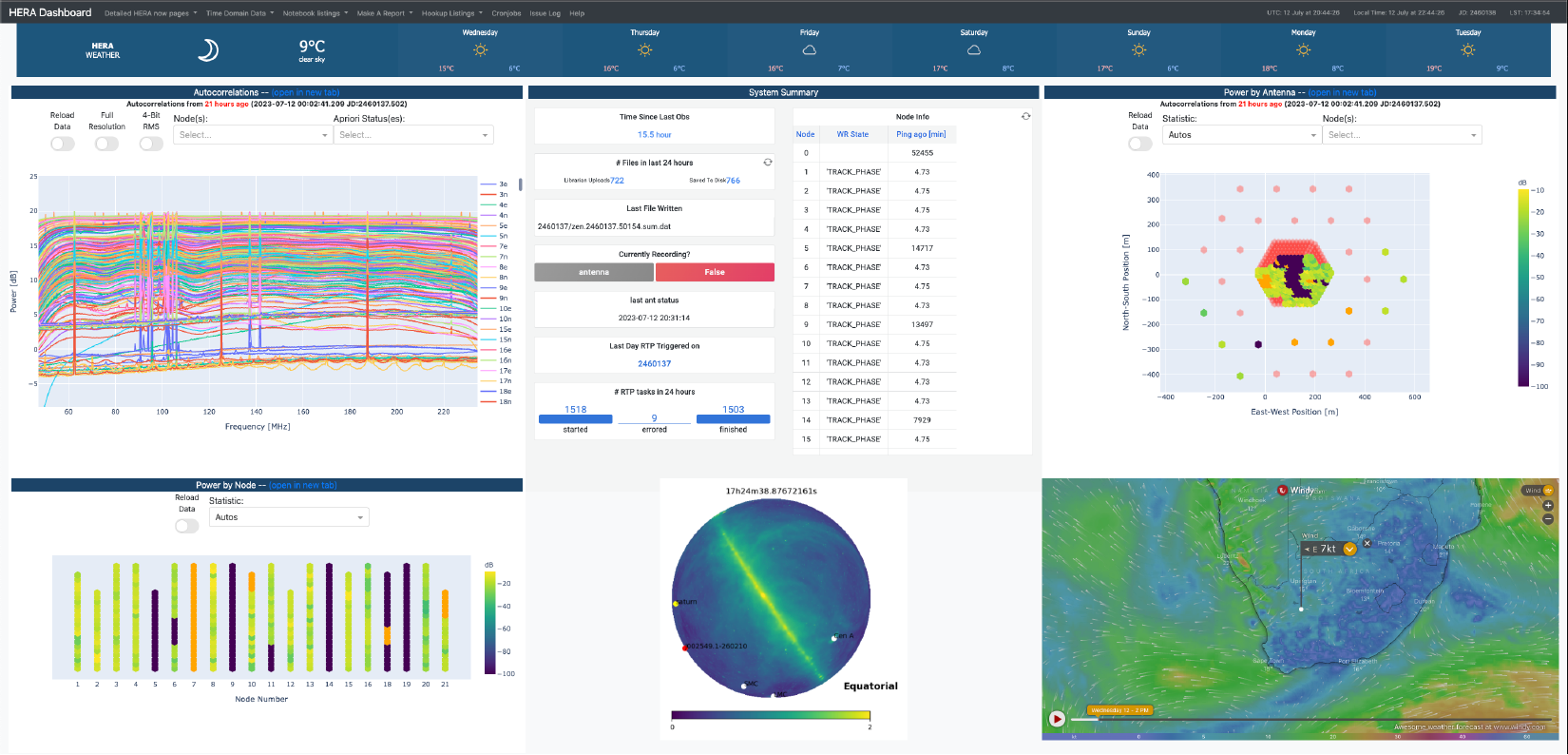}
    \caption{The home page of the dashboards hosted at heranow.reionization.org. The landing page offers a live plot of the autocorrelations from the array, a table of observing status, a median autocorrelation power plot in two views (by array position and node), major sources in the current sky above HERA, and a current weather widget. Observers can use this dashboard to make a quick assessment of the health of the array.}
    \label{fig:dashboard}
\end{figure*}

Web-based dashboards\footnote{heranow.reionization.org} provide a current and historical view of the telescope status derived largely from the M\&C database. The page includes live autocorrelation spectra, RF power levels organized by subsystem, and other relevant information needed to make a quick assessment about the overall array health. Figure \ref{fig:dashboard} shows a snapshot of the home page.

Additional dashboards provide a finer grained view into subsystems. Most use Grafana, where dashboard panes are constructed around an SQL query which can be further refined by user controlled variables like time range or node number. Figure \ref{fig:diskusage} showing RTP usage is one such example of a Grafana dashboard used by observers.

\subsubsection{Notebooks and Quality Metrics} \label{sec:notebooks}
Another part of observing is inspection of nightly analysis products. Several pipeline tasks produce Jupyter notebooks which are then automatically uploaded to a public web server. 

The notebooks include plots of raw data, interference detection, and calibration solutions. They also include full-day summaries of various array performance statistics. Several notebooks focus on antenna autocorrelations, which are a small, well understood dataset used to detect common system issues. Autocorrelation notebooks assess power levels that are too low (malfunctioning gain stage or broken fiber), too high (too much gain causing saturation) or abnormal passband for a spectrum, which may indicate a problem with the feed. Example cross-correlation metrics include  redundancy between repeated baselines and correlation of antennas \citep{Storer_2022}. These metrics are useful as automated functions to flag broken or misbehaving antennas, but they are also useful for observers to  identify unusual phenomena by eye. In practice, they have been used to identify a number of array issues, such as spurious correlations between antennas or timing failure identifiable as a loss of correlation. A few of these are discussed below in section \ref{sec:arrpef}. The RTP is increasingly comprised of Jupyter notebooks executed with \texttt{nbconvert}, allowing for simultaneous analysis and visualization. They serve as self-documenting logs, enabling quick debugging and identification of novel data issues.

\begin{figure*}
    \centering
    \includegraphics[width=2\columnwidth]{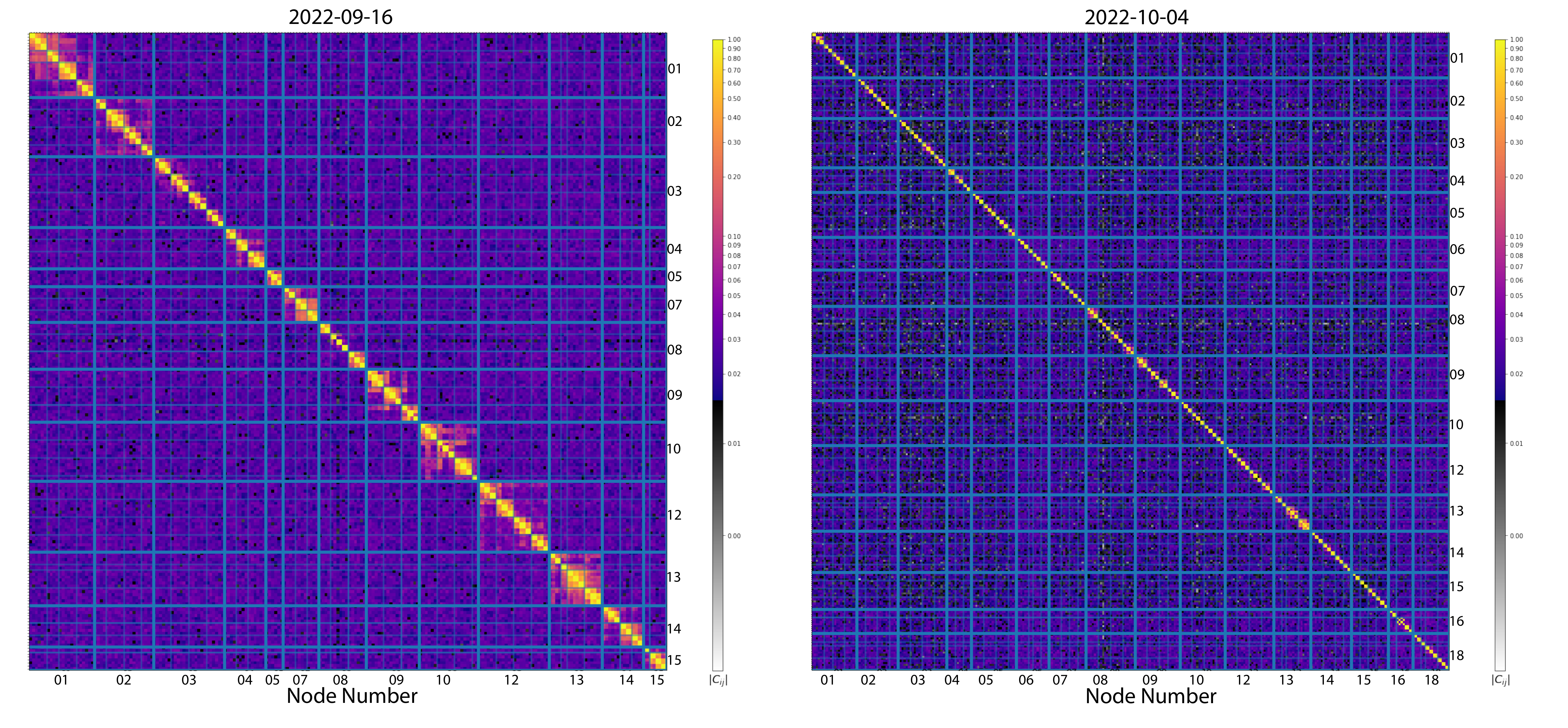}
    \caption{
    Correlation matricies illustrating the impact of improved signal chain isolation. This antenna-antenna correlation matrix is formed from visibilities normalized by auto correlations using antennas set to their internal loads. Each pixel represents a cross-correlation betwen two antennas in the array, meaning the diagonal is the autocorrelation. Correlation uses even/odd separated visibilities and should be mean-zero with a noise-like variance. The matrix was averaged over approximately 1 minute of time and is scaled such that noise has a black to grey scale and anything non-noise like is on a color scale. Before applying gaskets (left) a significant excess is seen across the array with largest values within a node. After applying gaskets (right) the structure within a node boundary is significantly reduced. Excess is still seen within the boundaries of a SNAP, indicating some remaining pathway for cross-coupling or common mode. }
    \label{fig:pamfix}
\end{figure*}

\section{Array Performance and Commissioning}\label{sec:arrpef}
HERA has operated continuously through a construction and test period spanning several years- as an array, we can operate and analyze sub-arrays of antennas as they come online. During this time, several systematics were identified in the phase II system which are worth describing in some detail because each degraded the overall performance, and fixing or mitigating the issue significantly improved the instrument. 
Here we largely show commissioning data, which is used to evaluate array performance, but does not constitute final science quality data products. A detailed description of the data from each season, the steps used to calibrate, flag, and all the other steps needed to obtain design sensitivity are left to future phase II results. For a description of these steps for the phase I system, see \citealp{phaseIlimits} and \citealp{heraresults2}. 

Systematics are a general category of unexpected instrument or analysis effects that degrade the data quality. To detect the EoR, the HERA telescope and analysis must be able to separate the EoR signal from from foregrounds which are far brighter than the EoR signal. Systematics that mix the bright foregrounds into regions of parameter space that HERA uses to detect the EoR, described in \ref{sec:pspec}, are of particular concern. Examples include unmodeled gain variation and digital non-linearity. All published deep 21 cm limits contend with one or more such behavior through a custom analysis which filters, flags, or calibrates away the unwanted effect, usually at the cost of additional simulation analysis to understand the impact on the resulting power spectrum (e.g, \citealp{phaseIlimits}, \citealp{heraresults2}, \citealp{lofarsystematic}, \citealp{Barry_2019}). HERA was built to a reference specification designed to avoid known systematics, but in practice subtle unexpected effects remain.  HERA's early observing program has presented the opportunity to find and correct several issues. Here we detail those which have had the largest impact on sensitivity.

\begin{figure*}
    \centering
    \includegraphics[width=1.8\columnwidth]{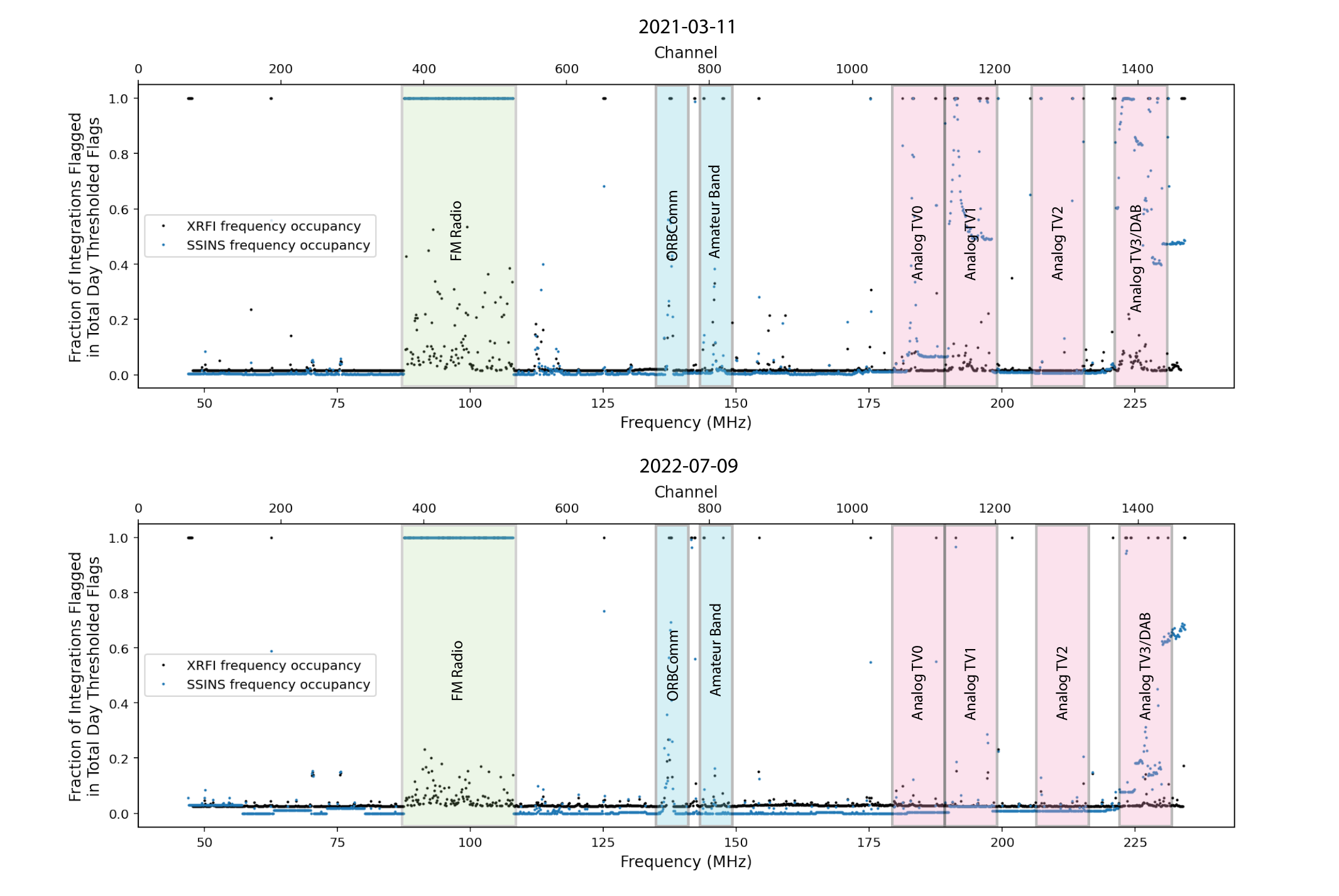}
    \caption{Single night RFI occupancy for two nights of HERA observing occuring before and after analog TV stations were turned off March 2021 and then in July 2022. Two sets of flags are produced from autocorrelations, one from XRFI (a simplified version of the flagger used for \citealp{phaseIlimits} and \citealp{heraresults2})  and one from SSINS (a parallel flagger used for commissioning). The most notable transmitter bands are labelled. In the earlier observing night (top), the Analog TV bands are still active and heavily flagged. In the later night (bottom), the emission is not evident. The Analog TV3 band overlaps with the Terrestrial Digital Audio Band (T-DAB), which is still evident in the later observing seasons. The FM band, ORBComm, and Amateur bands are also identified.}
    \label{fig:TV}
\end{figure*}

\subsection{Common Mode} \label{sec:crosstalk}
A prevalent issue, particularly in dense arrays like HERA, is the appearance of unwanted correlation between antennas (e.g, \citealp{Kern_2019}, \citealp{kwak2023effects}, \citealp{lofarmutcoup}, \citealp{josaitis_2022}). This issue takes many forms, but can generally be reduced into two categories, which in this paper will be called ``common mode'' and ``cross talk''. Common mode occurs when some unwanted signal leaks into two or more signal chains resulting in an excess correlation with a time dependence that is independent of sky fringes. This effect can possibly be ameliorated with better signal chain isolation, a phase switching system, or by temporally filtering the non-sky like components \citep{Kern_2019, Kern_2020}. Meanwhile, crosstalk happens when sky signal is unintentionally communicated from one antenna to another through antenna to antenna mutual coupling (e.g, \citealp{josaitis_2022}) or leakage between parts of the RF system \citep{heraresults2}.

In early HERA phase II observations, visibilities were observed to exhibit correlation beyond the expected level. This was particularly strong between antennas within the same node, and consequently within the same PAM receiver rack. The correlation is easily seen when plotting the visibility matrix normalized by the auto correlations \citep{Storer_2022}. This excess varied in time much faster than expected variation due to the sky suggesting another source of common mode or variation in the cross coupling path. It also, tellingly, was clearly visible even when antenna LNAs were turned off, suggesting a common source of RF power. The systematic was tracked in the lab to a gap in the aluminum cases enclosing individual PAM receiver cards. This gap allowed a pathway for common-mode signals to enter the system. It was demonstrated that it could be fixed by installing conductive foam gasket in the gap. Once applied to the field system, excess correlation seen within a node was significantly reduced. Figure \ref{fig:pamfix} illustrates the improvement with correlation matricies recorded with antenna switch state set to the 50 ohm internal load before and after application of gaskets around the RFoF cards. Each pixel represents a cross-correlation between two antennas in the array, meaning the diagonal is the autocorrelation. The antennas are sectioned by node. Excess correlation above the noise is visible across the array with much larger excess occurring within nodes. After the application of gaskets within all 350 PAM receivers excess correlation is significantly reduced across all visibilites and is now much closer to the noise level.  The anomalous temporal structure is also significantly reduced.

\subsection{Environmental RF Interference} \label{sec:rfi}
One of the most confounding systematics for radio instruments is radio frequency interference (RFI). Interference can come from a large variety of sources,  both man-made and natural. The Karoo site was chosen due to the relatively quiet RFI environment. However with satellites, planes, and meteors and other modes for over horizon propagation, no place on Earth will be entirely free of artificial RFI. Furthermore, there are sources that cannot be mitigated by moving away from them, such as natural phenomena and self-interference.

For man-made transmitters, the duty cycle of the transmitters can be quite high, causing the corresponding channels to be flagged out of the data at a very high percentage.  Fortunately, many of the most egregious transmitters are well documented and can be identified easily in the data. Typical emitters seen at HERA include the ORBCOMM satellite constellations at 137 MHz and FM radio at 88-110 MHz. Other times intermittent transmission path changes or use patterns can cause things to appear sporadically. 

\begin{figure}
    \centering
    \includegraphics[width=\columnwidth]{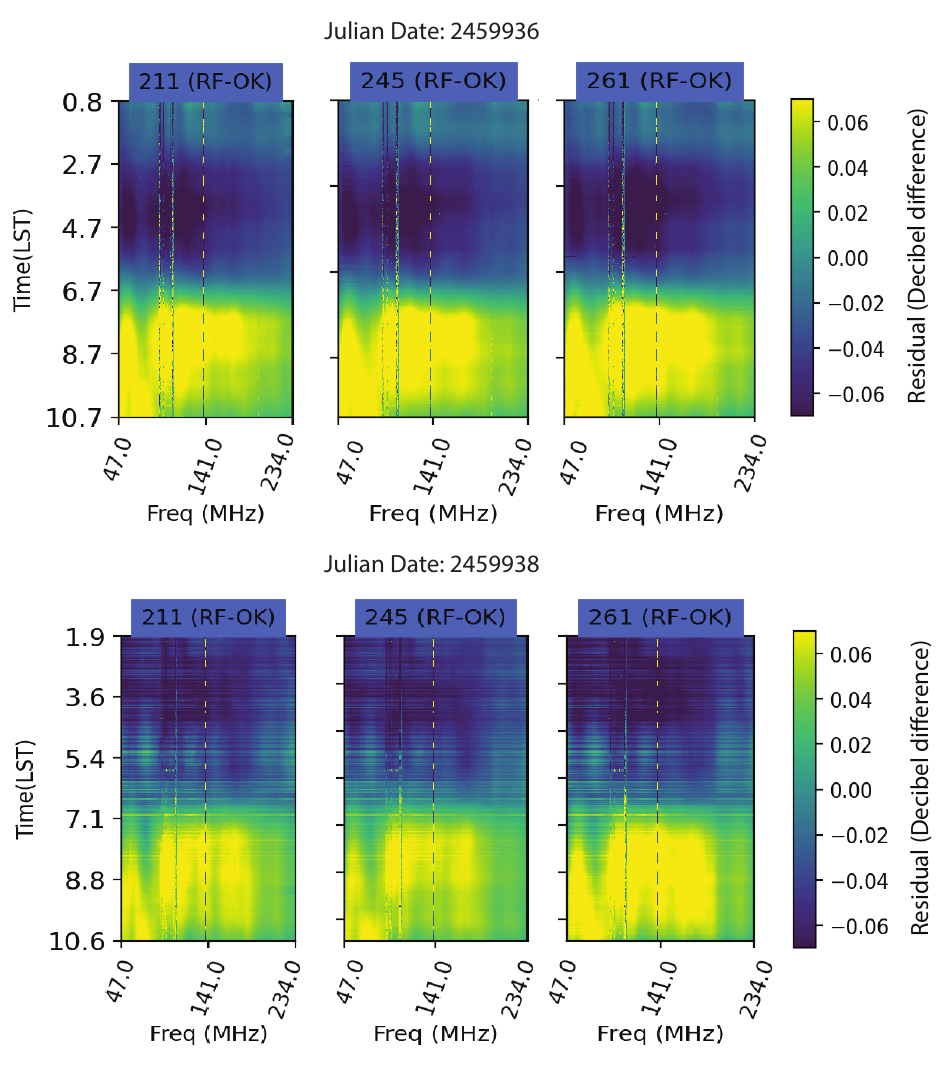}
    \caption{Autocorrelations on selected antennas illustrating the presence of a broadband signal with fast temporal structure (bottom) and a reference night (top) for comparison. A median has been subtracted from each channel to show variation. The temporal structure in the lower plot is likely caused by lightning. The ``RF-OK'' status of each antenna included in the subplot title indicates no known issues in the RF system.}
    \label{fig:lightning}
\end{figure}

\begin{figure*}[!ht]
    \centering
    \includegraphics[width=2\columnwidth]{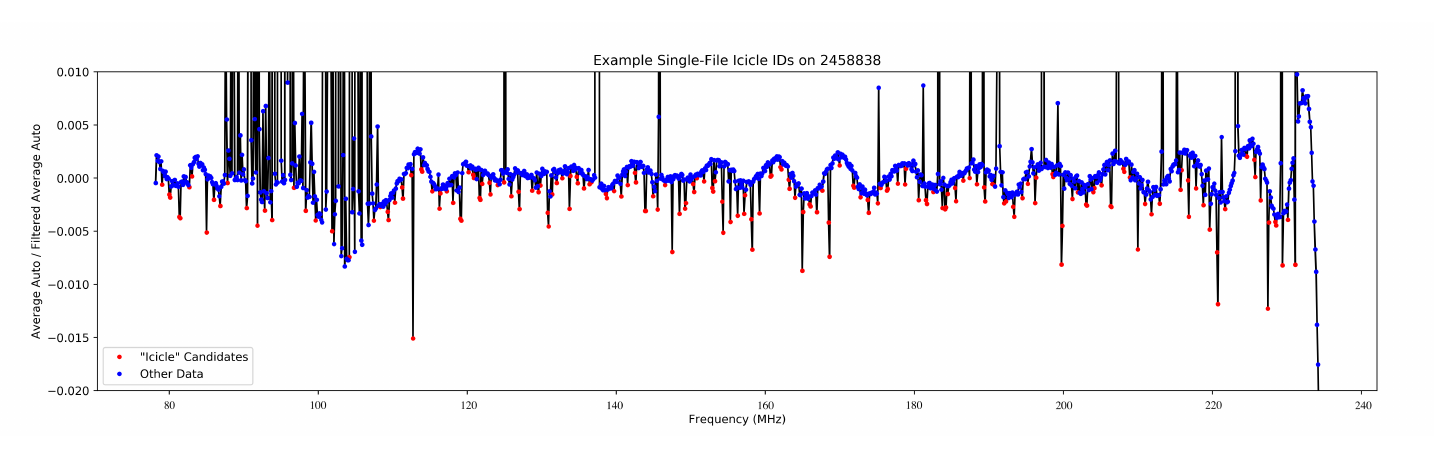}
    \caption{ Periodic negative icicles, highlighted with red dots, are seen in typical auto correlation spectrum after subtraction of a delay-filtered spectrum to highlight deviations at one part in 10,000. They are also visible in the cross-correlations but at lower levels. The problem was remedied by changing the FFT rounding scheme from round-to-infinity to round-to-even and alteration of the FFT-shift schedule to reduce quantization errors.}
    \label{fig:icicles}
\end{figure*}

Prior to the switch from analog to digital TV stations in South Africa, a few analog TV channels were present at high occupancy in HERA data.  Analog TV channels are allocated 6MHz of bandwidth and have a distinctive double pair of carriers separated by 4.5MHz with the lower frequency carrier usually about 10dB brighter.  At the time of the observation, digital TV was being deployed across South Africa and analog stations were being switched off. 
In the most recent observing season, these transmitters are no longer observed. In figure \ref{fig:TV} we illustrate this improvement comparing flags observed in March 2021 and then in July 2022 after the switch to digital. The flags were calculated with two methods: the Sky Subtracted Incoherent Noise Spectrum (SSINS) flagger \citep{ssins} and XRFI, the mainline HERA flagger at the time \citep{phaseIlimits, heraresults2}. These flags here are used for commissioning purposes and do not constitute flags used for final data products. The analog TV emission is more obvious in the the night taken before the switch to digital TV (top) and is no longer present in the later observations (bottom). FM radio and ORBCOMM are strongly present in both, and transmissions in the amateur band is sporadic throughout. The differing baselines between XRFI and SSINS, and the step-like structure in the baseline SSINS flags, are due to different flagging strategies. SSINS relies on user-defined shape templates based on the characteristics of the local spectrum allocation for flagging, which results in wider blocks for flags.

While most sources of RFI are relatively narrow-band, some nights have included incidences of signals that cause power over the entire bandwidth. This gain variation is often seen over the entire array and not on a per-antenna basis.  
Figure \ref{fig:lightning} illustrates such a situation compared to a more typical night. Snapshot imaging with a broad spectral range combined with information from the South African weather service suggests the source of this interference is usually lightning.  We estimate that about 10\% of the observing nights were affected. There were nights with high levels of broadband RFI that were not seen to be consistent with reported lightning, and this still needs to be investigated. For instance, out-of-band transmitters that cause saturation can cause a similar pathology.

\subsection{Signal Chain Linearity}
A multi-stage system has many potential saturation points in the analog or digital systems, which can cause the output visibilites to become non-linear with respect to the input RF sky power. Examples include amplifier compression in the LNA and numerical saturation during the FFT.

The most sensitive analog components are the low noise amplifiers and analog to digital converters. The symptoms of non-linearity at these points are a saturation of the spectrum at at least one channel and large ripples across the rest of the spectrum. As discussed in the previous section, RFI is often a culprit in signal chain compression. Analog components are designed for specific ranges of input power, and signals above the recommended input power levels can send the components out of their allotted range, causing unrecoverable signal distortion. 

There are built in mechanisms to assist with saturation issues, such as the adjustable attenuators built into the PAM receivers. Here 14 dB are available in 1dB steps to adjust a saturated antenna back into an allowable range for the ADC. However, should the saturation happen in the antenna, or should the power be more than 14 dB out of range the signal is unrecoverable. Therefore, environmental power level monitoring becomes very important. 

The digital system can also have non-linearity issues. Digital overflow can occur when an operation results in a number that is larger than the bit depth allocated for it can support. Overflow detection is built into the digital systems to monitor this, and designers make choices about how to handle overflows, such as whether to wrap (allow overflow, ignoring most significant bits of the signal), or clip (saturate at maximum value). Other sources of digital error may come from the inherent fixed precision of digital systems. Quantization of analog signals must be done in order to digitize them, and this comes with rounding errors that cause signal distortion at some level. These errors can be mitigated to some extend using methods such as the ``Van Vleck'' correction (e.g,\citealp{vanvleck}, \citealp{benkevitch2016van}). Reducing the bit number of the digitized data, which happens most notably when the output of the FFT is truncated from 8 to 4 bits, requires choices to be made about how to handle digits beyond the allowed binary representation for a particular operation, such as truncation or rounding. These re-quantizations are monitored to count overflows.

One example of a quantization issue that occurred in the HERA F-engine is shown in figure \ref{fig:icicles}, appearing as a periodic negative ``spike'' in the visibilities. The team labelled these ``icicles'' due to their distinctive downward shape. The digital system was identified as a likely culprit due to the negative bias of the systematic--real world RFI and analog distortion are expected to appear as positive power--and the regular spacing of the drops. The issue was isolated to a choice of rounding method in the FFT butterfly stages. After changing from round-to-infinity to round-to-even, and adjusting the FFT-shift schedule to reduce quantization errors, the icicles no longer appeared in data.

\begin{figure*}
    \centering
\includegraphics[width=1.8\columnwidth]{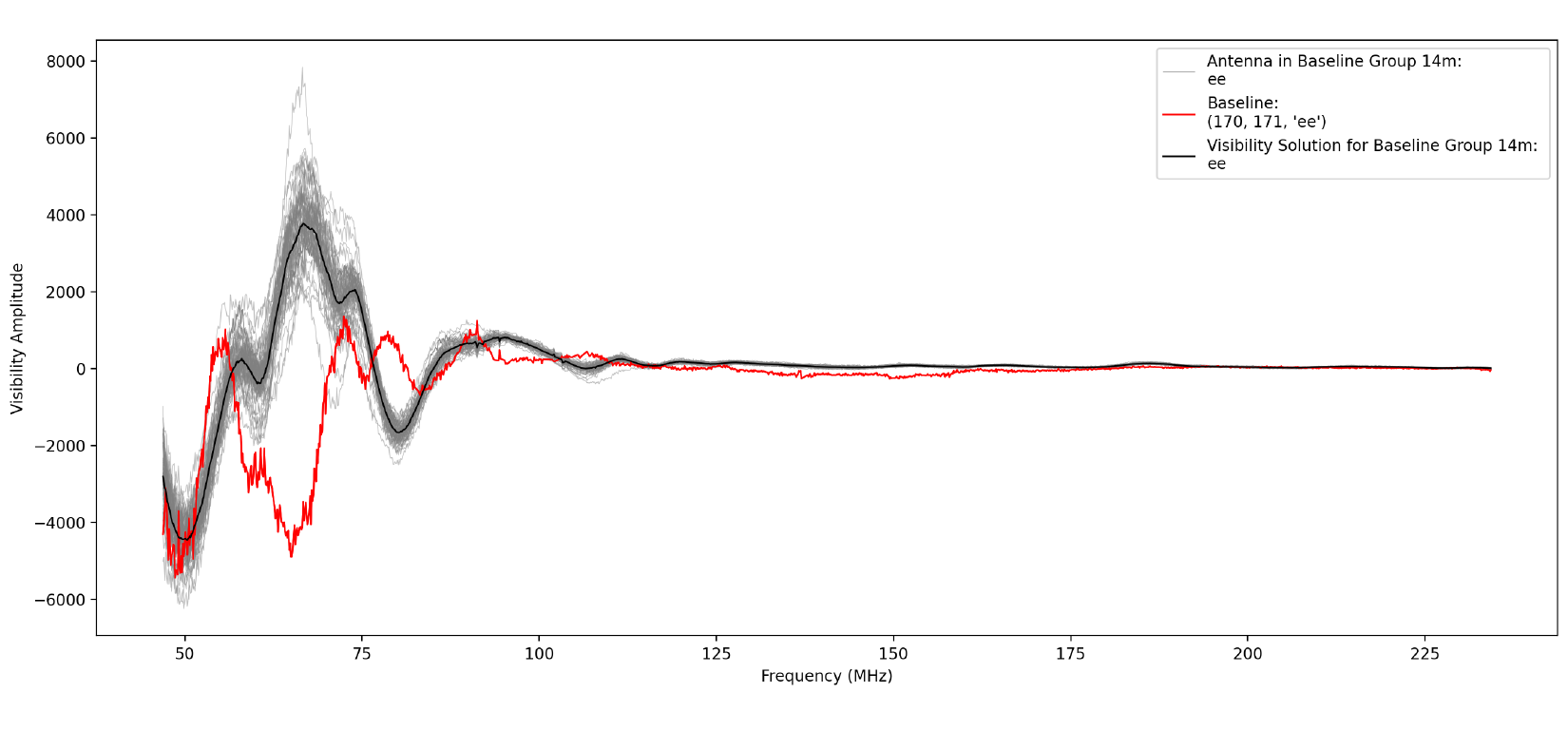}
    \caption{Redundant visibility spectra all simultaneously sampling the 14-m East-West oriented baseline. The black line shows the linear algebra visibility solution for the group, with individual baselines in gray. The highlighted red baseline contains an antenna that is a significant outlier in redundancy, but has no other metrics flagged. Typical chi-squared per antenna range from 0-3 for good antennas, and values above this are considered suspect. Antenna 171e reports a normal chi-squared for this file of 2.27, and 170e reports a statistical outlier at 20.}
    \label{fig:redundancy}
\end{figure*}

\subsection{Non-Redundancy}
A useful feature of the regular array antenna layout is using array redundancy for calibration and averaging. The core HERA antennas are arranged into a densely packed hexagonal lattice broken into three offset subarrays. 

Though sensitivity was the primary driver for this choice, calibration was also a factor. When HERA was designed, errors in sky model were already known to introduce chromaticity \citep{datta2010, barry2016, Ewall_Wice_2017,byrne2019} while knowledge about the low frequency sky was still incomplete. 
With antennas arranged in a repeating lattice per-antenna gains can be solved for with linear algebra \citep{Liu_2010, Zheng_2014, Dillon_2020}, though this is likewise vulnerable to introducing chromaticity when the array is not sufficiently redundant \citep{Orosz_2019}. Furthermore, a sky model is still necessary to solve for the final few degrees of freedom, like the absolute flux scale \citep{Dillon_2018, Kern_2020b}. This step has also been shown to introduce unwanted chromaticity in the presence of sky model errors \citep{byrne2019} but this can potentially be overcome with a unified process \citep{byrne2021} or the addition of the autocorrelations \citep{li2019}. HERA's primary pipeline avoids adding additional spectral structure due to calibration errors by smoothing gain spectra on cosmologically relevant frequency scales \citep{Kern_2020b}, at the risk of not calibrating out any real fine-scale spectral structure in the signal chains. Lastly we note that averaging across sets of redundant baselines significantly reduces data volume to speed up post-calibration processing. 

In its simplest form, the redundant calibration technique assumes that the system can be modeled with single complex gain for each antenna and that all antennas are actually on their grid locations. The resulting solution from redundant calibration is a set of smoothed gain spectra for each antenna, a single consensus visibility solution for each baseline, and a chi-square value for each antenna. In reality, baselines are not perfectly redundant. Deviations in antenna position or beam pattern can contribute to baselines being less-than-perfect copies of one another and degrade the quality of the calibration solution. Figure \ref{fig:redundancy} shows visibility spectra for a set of calibrated redundant baselines to give some idea of typical variation after calibration. The red highlighted baseline contains one nominally good antenna (171e) and one antenna that is a significant outlier in redundancy (170e). Typical chi-squared per antenna range from 0-3 for good antennas, and values above this are considered suspect.  Antenna 171e reports a normal chi-squared for this file of 2.27, and 170e is a significant outlier at 20. Antenna 170e was not flagged for any autocorrelation metrics but is flagged out of the data after this step.

The reason for non-redundancy is not identified for this antenna. Redundancy metrics have been used to identify and correct issues such as swapped feed polarizations and incorrect feed height. Though other individual causes of non-redundancy remain unsolved, the array average has been seen to improve after regular maintenance suggesting that non-redundancy is a useful indicator of overall array health.

\section{Conclusion}
In recent years, HERA has upgraded from the phase I commissioning array to the phase II system as described in this paper. Results from the phase I system have been published and represent some of the most sensitive EoR limits to date. Results from the phase II system will improve even further upon previous results.  The phase II system upgrades the array to cover the full planned 50-250 MHz frequency band, extending the array's redshift range to cosmic dawn observations. At the time of writing, the phase II system has completed two seasons of observing and more are planned.

\section*{Acknowledgments}
This material is based upon work supported by the National Science Foundation under Grant Nos. 1636646 and 1836019 and institutional support from the HERA collaboration partners. This research is funded by the Gordon and Betty Moore Foundation through
grant GBMF5215 to the Massachusetts Institute of Technology. HERA is hosted by the South African Radio Astronomy Observatory, which is a facility of the National Research Foundation, an agency of the Department of Science and Innovation. LMB acknowledges that this material is based upon work supported by the National Science Foundation Graduate Research Fellowship under Grant No. 2233001. JSD gratefully acknowledges the support of the NSF AAPF award \#1701536. A. Liu acknowledges support from the New Frontiers in Research Fund Exploration grant program, the Canadian Institute for Advanced Research (CIFAR) Azrieli Global Scholars program, a Natural Sciences and Engineering Research Council of Canada (NSERC) Discovery Grant and a Discovery Launch Supplement, the Sloan Research Fellowship, and the William Dawson Scholarship at McGill. Y.-Z. Ma acknowledges the support of the National Research Foundation of South Africa with Grant No.~150580. D. Storer acknowledges that this material is based upon work supported by the National Science Foundation Graduate Research Fellowship Program under Grant No. DGE-1762114. Any opinions, findings, and conclusions expressed in this material are those of the author(s) and do not necessarily reflect the views of the National Science Foundation. MGS and PK acknowledge support from the South African Radio Astronomy Observatory (SARAO) and the National Research Foundation (Grant No. 84156). G.B. acknowledges support from the Ministero degli Affari Esteri della Cooperazione Internazionale - Direzione Generale per la Promozione del Sistema Paese Progetto di Grande Rilevanza ZA18GR02 and the National Research Foundation of South Africa (Grant Number 113121) as part of the ISARP RADIOSKY2020 Joint Research Scheme. NSK acknowledges support from NASA through the NASA Hubble Fellowship grant \#HST-HF2-51533.001-A awarded by the Space Telescope Science Institute, which is operated by the Association of Universities for Research in Astronomy, Incorporated, under NASA contract NAS5-26555. EdLA acknowledges the financial support of STFC.

\bibliography{PASPsample631.bib}
\bibliographystyle{aasjournal}

\end{document}